\documentclass[12pt,a4paper,final]{iopart}
\pdfoutput=1
\usepackage{amscd,amssymb}
\usepackage{epsfig}
\usepackage{amscd,amssymb}
\usepackage{graphicx,xspace}
\usepackage{pstricks}
\usepackage[left=3cm,top=3cm,right=3cm,bottom=3cm]{geometry}
\usepackage{cite}
\usepackage{bigfoot}
\usepackage{appendix}
\usepackage[breaklinks=true,colorlinks=true,linkcolor=black,urlcolor=black,citecolor=black,pdfencoding=auto]{hyperref}
\usepackage{amsfonts, amsmath, dsfont}
\usepackage{empheq}
\usepackage{bookmark}
\usepackage{color}
\definecolor{labelkey}{cmyk}{.4,.2,0,0}

\usepackage{ulem}

  \usepackage{verbatim}

  \definecolor{blue}{rgb}{0,0,1}
  \definecolor{green}{rgb}{0,.6,0}
  \definecolor{red}{rgb}{1,0,0}
  \definecolor{vio}{rgb}{1,0,1}
  \definecolor{uv}{rgb}{0.5,0,0.5}
  \definecolor{ama}{rgb}{0.3,0.3,0.3}

\renewcommand{\leq}{\leqslant}
\renewcommand{\geq}{\geqslant}

\newcommand \be  {\begin{equation}}
\newcommand \bea {\begin{eqnarray} }
\newcommand \ee  {\end{equation}}
\newcommand \eea {\end{eqnarray}}

\newcommand{\ket}[1]{|\kern.3ex#1\kern.3ex\rangle}
\newcommand{\bra}[1]{\langle\kern.3ex #1 \kern.3ex|}

\begin{document}

\title[Hessian at the minimum for manifolds in a high-dimensional random landscape]{Manifolds pinned by a high-dimensional random landscape: Hessian at the global energy minimum.}
\vskip 0.2cm

%
%
%
%
%
%
%
%

\author{Yan V. Fyodorov}
\address{King's College London, Department of Mathematics, London  WC2R 2LS, United Kingdom}
\author{Pierre Le Doussal}
\address{Laboratoire de Physique de l'Ecole Normale Sup\'erieure,
PSL University, CNRS, Sorbonne Universit\'es,
24 rue Lhomond, 75231 Paris, France}

\date{\today}

\begin{abstract}
We consider an elastic manifold of internal dimension $d$ and length $L$ pinned in a $N$ dimensional random potential
and confined by an additional parabolic potential of
curvature $\mu$. We are interested in the mean spectral density $\rho(\lambda)$ of the Hessian matrix $K$ at the absolute minimum of
the total energy.
We use the replica approach to derive the system of equations for $\rho(\lambda)$ for a fixed $L^d$ in the $N \to \infty$ limit extending 
 $d=0$ results of our previous work \cite{UsToy}.
 A particular attention is devoted to
analyzing the limit of extended lattice systems by
letting $L\to \infty$. In all cases
we show that for a confinement curvature $\mu$ exceeding a critical value $\mu_c$,  the so-called "Larkin mass", the system is replica-symmetric and
 the Hessian spectrum is always gapped (from zero). The gap vanishes quadratically at $\mu\to \mu_c$.  For $\mu<\mu_c$  the replica symmetry breaking (RSB) occurs and the Hessian spectrum is either gapped or extends down to zero,
depending on whether  RSB is 1-step or full. In the 1-RSB case the gap vanishes in all $d$
as $(\mu_c-\mu)^4$ near the transition.
In the full RSB case the gap is identically zero. A set of specific landscapes realize
the so-called "marginal cases" in $d=1,2$ which share both feature of the 1-step and the full RSB solution,
and exhibit some scale invariance. We also obtain the average Green function
associated to the Hessian and find that at the edge of the spectrum it decays exponentially in the distance 
within the internal space of the manifold with a length scale equal in all cases to the 
Larkin length introduced in the theory of pinning.
\end{abstract}

\maketitle


\section{Introduction}

\subsection{The random manifold model and some known results}

Numerous physical systems can be modeled by a collection of points or particles coupled by
an elastic energy, usually called an elastic manifold, submitted to a random potential
(see \cite{BlaFeiGesLarVin94,LeD11,TGPLDBragg2} for reviews).
They are often called "disordered elastic systems" and generically exhibit pinning in their statics
and depinning transitions and avalanches in their driven dynamics
\cite{Fis85,Nattermann1992,LeDWieCha02,RosKra02,RosLeDWie09,LeDWie13}.
Their energy landscape is complex leading to glassy behavior.

The manifold is usually parameterized by a $N$-component real displacement field ${\bf u}(x) \in \mathbb{R}^N$,
where $x$ belongs to an internal space $x \in \Omega$. $\Omega$ can be either a finite collection
of points, such as a subset $L^d$ of an internal space of dimension $d$,
$\Omega \subset \mathbb{Z}^d$, for discrete models,
or $\Omega \subset  \mathbb{R}^d$ in a continous setting. The case $d=1$ corresponds to a line in $N$ dimensions and for $N=1$ was studied in the present
context in \cite{FLRT}. The case $d=0$ usually
refers below to $\Omega$ being a single point, previously studied in \cite{UsToy} in the large $N$ limit, and
the present study can be seen as its generalization to a manifold. There are
two terms in the total energy. First the points in $\Omega$ are coupled via an elastic energy, which is a quadratic form in the fields ${\bf u}(x)$. We also include in this quadratic term a parabolic confining potential
of curvature $\mu>0$.
The absolute minimum of this first term is thus the flat, undisturbed, configuration ${\bf u}(x)=0$.
The second term is the quenched disorder, modeled by a random potential energy which couples directly to ${\bf u}(x)$. We thus consider the following model of an elastic manifold in a random potential given by
its energy functional
\begin{equation}\label{landscape}
{\cal H}[{\bf u}]=\sum_{x,y} {\bf u}(x) \cdot (\mu \mathbf{1} - t \Delta)_{xy} \cdot {\bf u}(y)
+ \sum_{x} V(\mathbf{u}(x),x)
\end{equation}
where here $x \in L^d \subset \mathbb{Z}^d$, $\mathbf{1}$ is an appropriate identity operator, and the
matrix $- t \Delta_{xy}$ is required to be positive definite.
Here $\Delta$ can be chosen as the discrete Laplacian in the hypercube $L^d$ with periodic boundary conditions. In that case its eigenmodes are plane waves $\sim e^{i k x}$ and we denote $\Delta(k)$ its eigenvalues, i.e. in $d=1$, $\Delta(k)= 2 (\cos k-1)$ with $k=2 \pi n/L$, $n=0,..L-1$. For general $d$ similar formula holds and $t$ must be
positive, $t>0$.
All formula below extend immediately to more general functions $t \Delta(k)$, e.g. to more general elasticity
(such as long range elasticity). They also extend to cases where $t \Delta_{xy}$
is a quadratic form defined on any graph $\Omega$. Finally, they also extend to the
limit of the continuum manifold model, e.g. with the standard Laplacian $\Delta=\sum_{i=1}^d \frac{\partial^2}{\partial x_i^2}$  whose spectrum is given by $\Delta({\bf k})= -{\bf k}^2$. We thus use the notation $\int_k = \frac{1}{L^d} \sum_k  \equiv \int \frac{d^d k}{(2 \pi)^d}$ so our main formula are valid both for discrete and continuum (in the continuum $\sum_x \equiv \int d^dx$).
 We see from \eqref{landscape} that $\mu$ acts as a ``mass" which, for the continuum model, leads to reducing the fluctuations
beyond the scale $L_\mu= \sqrt{t/\mu}$.

Here we will consider $V({\bf u},x)$ to be a mean-zero Gaussian-distributed random potential in $\mathbb{R}^N \times
\mathbb{Z}^d$
with a rotational and translational invariant covariance (also called the {\it correlator} in the physical context)
such that potential values are uncorrelated for different points in the internal space, but correlated for different displacements:
\footnote{We follow here the
same notations as in \cite{MP2,TGPLDBragg1a,PLDKWLargeNDetails,LDMW}.
Note the factor of $2$ difference with the definition of $B(q)$ in \cite{UsToy}.
Noting $\hat B$ the similar function there, we have $B(q)=\hat B(q/2)$.
$Q$ defined below is the same object in both papers.}
\begin{equation}\label{cov}
\overline{ V(\mathbf{u}_1,x_1) V(\mathbf{u}_2,x_2) } = N\: B\left(\frac{(\mathbf{u}_1-\mathbf{u}_2)^2}{N}\right) \delta^d(x_1-x_2),
\end{equation}
In Eq.(\ref{cov}) and henceforth the notation
$\overline{ \cdots}$ stands for the quantities
averaged over the random potential.

The equilibrium statics of this model has been much studied. From the competition
between the elastic and the disorder energy, the minimal energy configuration ${\bf u}_0(x)$ (ground state) is non trivial and exhibits interesting statistically self-affine properties characterized by a roughness exponent:
${\bf u_0}(x) - {\bf u_0}(0) \sim |x|^\zeta$. The
sample to sample fluctuations of the ground state energy (and, at finite temperature, of the free energy)
grow with the scale as $\sim L^\theta$, with $\theta=d-2+2 \zeta$ as a consequence of the symmetries
of the model \eqref{landscape}-\eqref{cov}. In addition the manifold is {\it pinned}, i.e. its macroscopic response to an external force is non-linear. The early (and partly phenomenological) theory of pinning is due to Larkin and Ovchinnikov (see \cite{BlaFeiGesLarVin94} for a review).
Below the so-called Larkin length scale
$L_c$, with $L_c \sim (B''(0))^{-1/(4-d)}$ for weak disorder (small $B''(0)$) the deformations
are elastic, the response is linear, deformations can be calculated from perturbation theory,
leading to the roughness exponent $\zeta=(4-d)/2$. Above $L_c$ metastability sets in, the response to perturbations involves jumps (shocks), with non-trivial roughness $\zeta$ of minimal energy configurations.
Describing that regime has been a challenge, and progress was later achieved using the
bag of tools of the statistical mechanics of disordered systems, most notably replica methods.
Exact results have been obtained, but in only a few analytically tractable cases.
The first of such cases are mean field type models, notably the model \eqref{landscape} in the
limit $N \to \infty$. Saddle point equations in replica space \cite{MezPar91,BalBouMez96,LDMW,PLDKWLargeNDetails}
lead to solutions
exhibiting replica symmetry breaking (RSB) for $\mu < \mu_c$,
which describe the glass phase where the manifold is pinned.
The critical mass $\mu_c$ corresponds to the Larkin scale $L_c = \sqrt{t/\mu_c}$
and the glass phase appears at scales exceeding $L_c$. A second set of results were
obtained using the functional renormalization group
\cite{LeD11,Fis86,LeDWie04,LeD09AnnPhys} and are valid in an expansion
around $d=4 - \epsilon$ (for any $N$). While the resulting physical picture is
somewhat different, these could be reconciled \cite{PLDKWLargeNDetails,LDMW}.
Note also that the Larkin picture was fully confirmed by these studies.
Finally, for $d=1$ the problem can be mapped to stirred Burgers and
Kardar-Parisi-Zhang growth (see Ref.~\cite{HalZha95} for review of earlier works).
For $N=1$, a number of exact results were obtained recently from
an emerging integrability structure of the theory, both in physics and
mathematics. Besides proving the exact roughness $\zeta=2/3$ and free energy fluctuation exponent,
$\theta=1/3$, it was shown, e.g., that the probability density of the free energy for a long polymer
converges to the famous Tracy-Widom distribution both at
zero temperature \cite{Joh00}, and at finite temperature in the continuum
\cite{CalLeDRos10,Dot10a,Dot10b,SasSpo10,AmiCorQua11}.
Finally, note that the model \eqref{landscape}-\eqref{cov}
also arises in the study of the decaying Burgers equation with a random initial conditions in dimension $N$, which
exhibits interesting transitions and regimes, see e.g. for $N=1$ \cite{YanPLDBurgers}
and for large $N$ \cite{PLDMarkusBurgers}.

\subsection{Motivation and goals of the paper}

While these results predict large scale properties of the
low energy configurations, little is known about the detailed statistical structure of the complex
energy landscape of pinned manifolds. This relates to the broad effort of understanding the statistical structure of stationary points (minima, maxima and saddles) of random landscapes which is  of steady interest in theoretical physics~\cite{Lon60,HalLax66,WeiHal82,Fre95,AnnCavGiaPar03,Fyo04,Par05,BraDea07,FyoWil07,FyoNad12}, with recent applications to statistical physics \cite{AnnCavGiaPar03,Fyo04,Par05,FyoWil07,FyoNad12,FyoLeD14,FLRT,RBBC2018}, neural networks and complex dynamics \cite{WaiTou13,FyoKho16,Fyo16,Ipsen2017,IpsenForr2018}, string theory~\cite{DouShiZel04,DouShiZel06} and cosmology~\cite{EasGutMas16,YamVil2018}.
  It is also of active current interest in pure and applied mathematics~\cite{AzaWsc09,AufBenCer13,AufBen13,Nic14,Fyo15,SubZei15,Sub2017,Sub2018,CamWig17,ChSch},
 For the model \eqref{landscape}-\eqref{cov} in the simplest
case $d=0$ ($x$ is a single point), the mean number of stationary points and of minima of the
energy function was investigated in the limit of large $N \gg 1$ in \cite{Fyo04,FyoWil07,FyoNad12}, see also \cite{BraDea07,Fyo15,YamVil2018}. It was found that a sharp transition occurs from a 'simple' landscape for $\mu>\mu_{c}$ (the same $\mu_c$ as given by the onset of RSB, see above), with typically only a single stationary point (the minimum) to a complex ('glassy') landscapes for $\mu<\mu_c$ with exponentially many stationary points.
Similar transitions were found in related systems upon applying various external perturbations \cite{FyoLeD14,Fyo16,RBBC2018} in particular in the mean number of stationary points
which was also studied recently for the case of an elastic string $d=1$ in dimension $N=1$ \cite{FLRT}.
Relations with Anderson localization was discussed there in this context.

An important quantity which characterizes the stability of local equilibria, and is crucial
both for equilibrium and slowly driven dynamics, is the Hessian matrix. In particular,
the question of whether the spectrum of the Hessian at low lying local minima is
gapped (away from zero) or not, the behavior of its mean density of eigenvalues near
zero, and the nature of the associated low lying modes, has been identified as a
crucial feature to describe classical \cite{FPUZ,MuellerWyart14,MuellerWyart15}
and quantum glasses
\cite{AndMuel,TGPLDQuantum,LCTGPLDWigner,SchehrSpecificHeat,ChalkerGurarieBosons}.
Clearly, a 'gapless' spectrum reflects the existence of very 'flat' directions in configuration
space along which moving away from the local minimum incurs very little 'cost'.
This flatness, also known as a 'marginal stability', is ubiquitous in various types of
glasses \cite{MuellerWyart14,MuellerWyart15} and appears naturally in
models exhibiting a hierarchical structure of the energy landscapes
\cite{MPV,MLC2006}. The Hessian matrix was studied recently numerically in the context of the depinning
of an elastic line $d=1$ in a one dimension random potential, $N=1$, in an effort to
identify the "soft modes" which trigger the avalanches. It was found that in the stationary state
reached upon quasi-static driving, the low-lying modes of the Hessian are localized, with a localization length directly
related to the Larkin pinning length \cite{CaoRossoSoftModes}.
Although studying the Hessian at equilibrium, and specifically at the global minimum would be
also very interesting, it is analytically challenging for $d$ or $N$ small.

Recently, by combining methods of random matrix theory with methods of statistical mechanics of disordered systems, we were able to study the Hessian at the absolute minimum for the particle model ($d=0$)
in the limit of large $N \to \infty$ \cite{UsToy}. The main goal of the present paper is to
extend this study to the pinned elastic manifold. Hence we will study the
$N L^d \times N L^d$ Hessian matrix
\be \label{Hessian}
K_{ix,jy}[{\bf u}]=\frac{\partial ^2}{\partial u_{i}(x) \partial u_j(y)}{\cal H}[{\bf u}]
=\delta_{ij} (\mu \mathbf{1} - t \Delta)_{xy} + \delta_{xy} \frac{\partial ^2}{\partial u_{i}\partial u_j}V({\bf u}(x),x)
\ee
in particular its density of eigenvalues $\rho(\lambda)$ normalized as $\int \rho(\lambda)\,d\lambda=1$.
An important feature of such matrix is its (block-)band structure visualized below:

\begin{picture}(150,250)(-20,-60)
\put(151,-20){\thicklines\line(1,0){58}}
\put(-20,160){\thicklines \line(1,0){170}}
\put(-20,-19){\thicklines\line(0,1){179}}
\put(151,-79){\thicklines \line(0,1){119}}
\put(151,160){\thicklines \line(1,0){58}}
\put(-20,-80){\thicklines \line(1,0){229}}
\put(-20,-80){\thicklines \line(0,1){229}}
\put(209,-80){\thicklines \line(0,1){240}}
\put(25,10){\begin{picture}(20,20)

\multiput(-45,90)(57,-60){4}{\thicklines\line(1,0){56}}
\multiput(13,90)(57,-60){2}{\thicklines\line(1,0){56}}
\multiput(-45,90.5)(57,-60){3}{\thicklines\line(0,1){60}}
\multiput(12,90)(57,-60){2}{\thicklines\line(0,1){59}}
\put(-43,130){$(\mu+2t){\bf 1}_N$}
\put(135,115){$-t{\bf 1}_N$}
\put(-40,-60){$-t{\bf 1}_N$}
\put(-38,110){$+{\bf W}^{(1)}$}
\put(21,115){$-t{\bf 1}_N$}
\put(13,70){$(\mu+2t){\bf 1}_N$}
\put(23,50){$+{\bf W}^{(2)}$}

\put(-39,55){$-t{\bf 1}_N$}
\put(79,55){$-t{\bf 1}_N$}
\put(23,-6){$-t{\bf 1}_N$}
\put(79,18){\circle*{4}}
\put(86,10){\circle*{4}}

\put(111,-14){\circle*{4}}
\put(118,-22){\circle*{4}}
\put(78,-60){$-t{\bf 1}_N$}
\put(128,-55){$(\mu+2t){\bf 1}_N$}
\put(134,-71){$+{\bf W}^{(L)}$}
\put(134,-5){$-t{\bf 1}_N$}

\end{picture}}

\put(22,-100){$\mbox{\sf Structure of the} \,\, (NL\times NL) \,\, \mbox{\sf Hessian matrix } $K$\,\, \mbox{\sf in 1d discrete lattice }$}
\put(22,-120){$\mbox{\sf model with $L$ internal sites and periodic boundary conditions.}$}
\put(22,-140){$\mbox{\sf  Only non-zero $N\times N$ blocks are indicated.}$}
\label{banded}
\end{picture}

\vspace{4cm}

\noindent where for $r=1,\ldots,L$ we have introduced $N\times N$ random matrices ${\bf W}^{(r)}$
with entries  ${\bf W}^{(r)}_{ij}= \frac{\partial ^2}{\partial u_{i}\partial u_j}V({\bf u}(x),x)\big|_{x=x_r}$.

 Our main focus here is the problem where the Hessian $K_{ix,jy}[{\bf u}_0]$
is chosen at the global minimal energy configuration ${\bf u}_0\equiv {\bf u}_0(x)$.
At the same time it is worth noting another interesting problem, where
the Hessian is not conditioned by the global energy minimum, but instead chosen
at a generic point in configuration space, i.e. at an arbitrary {\it fixed}
${\bf u}(x)$. It is easy to see from \eqref{Hessian} and from the statistical translational
invariance of the correlator in (\ref{cov}) that
the Hessian is then statistically independent of the choice of ${\bf u}(x)$,
i.e. we may as well chose it at ${\bf u}(x)={\bf 0}$. The covariance structure of the random potential (\ref{cov}) implies, after a simple differentiation
 that entries of the matrices ${\bf W}^{(r)}$ are mean-zero Gaussian-distributed, independent for different $r$ and have the following covariance structure:
\begin{equation}\label{covWdisc}
 \left\langle  {\bf W}^{(r)}_{ij}{\bf W}^{(s)}_{kl} \right\rangle  =\delta_{rs}\frac{4}{N}B''(0)\left(\delta_{ij}\delta_{lk}+\delta_{ik}\delta_{jl}+\delta_{il}\delta_{jk} \right)
\end{equation}
The matrices of such block-band type, with ${\bf W}^{(r)}$ in diagonal blocks replaced with GOE matrices with i.i.d. entries, were introduced by Wegner\cite{Wegner1979} in his famous studies of the Anderson localization,  and are now  known by the general name of Wigner orbital models.
Various instances of the models kept attracting attention in Theoretical and Mathematical Physics literature over the years,
see e.g. recent paper \cite{Peled_etal} and references therein.
In particular, the mean eigenvalue density for such type of models as $N\to \infty$ is known to be determined by the {\it deformed semicircle} equation rigorously derived in \cite{KhorPast1993}. That equation naturally generalizes the so-called Pastur equation of random matrix theory  \cite{Pastur}. We will see below that the difference between GOE covariance and our choice (\ref{covWdisc}) is immaterial for the calculation of the mean eigenvalue density which will be found to satisfy exactly the same deformed semicircle equation. Moreover, when we condition the Hessian by being at the global energy minimum the equation retains its validity, albeit with the renormalized
curvature parameter $\mu\to \mu_{eff}$, which should be determined by a separate minimization procedure.
 The replacement $\mu\to \mu_{eff}$ is crucial in determining the {\it global position} of the support of the density of states, i.e. the position of the edge(s) and the value of the gap, for the Hessian at the global energy minimum, but
the general form of the density can be already determined without that knowledge
by studying the above mentioned equation.

An analysis of the density profile following from that equation hence comes naturally in our problem as by-product of its solution.
Surprisingly, we were not able to trace it in the literature for the most interesting case of infinite system of size $L\to \infty$.
As it may have a separate interest and is quite instructive, we are going to fill in that gap in the present paper and provide such an analysis for $d=1$.

In the case of a continuous manifold the Hessian matrix $K$ becomes a matrix-valued differential operator ${\cal K}$ acting in the space of $N-$component vectors ${\bf f}(x):=\left(f_1(x),\ldots,f_N(x)\right)^T$ where, e.g. $x\in [0,L]^d$, by the following rule:
\be \label{Hessianoper}
{\cal K}{\bf f}=(\mu \mathbf{1} - t \Delta){\bf f}+\hat{W}{\bf f}, \quad W_{i,j}(x)=\frac{\partial ^2}{\partial u_{i}\partial u_j}V({\bf u}(x),x)
\ee
with appropriate boundary conditions (e.g. periodic, or Dirichlet).

Without conditioning by global minimum the covariance structure of $\hat{W}$ is a natural analogue of (\ref{covWdisc}):
\begin{equation}\label{covWcont}
 \left\langle   W_{i,j}(x_1) W_{k,l}(x_2) \right\rangle  =\delta(x_1-x_2)\frac{4}{N}B''(0)\left(\delta_{ij}\delta_{lk}+\delta_{ik}\delta_{jl}+\delta_{il}\delta_{jk} \right)
\end{equation}

In particular, for $d=1$ the operator can be visualized in the following form of an $N\times N$ matrix
\footnote{Note that in such a case the spectral density in the infinite-volume limit can not be normalized. E.g. recall the density for the disorder-free model with $N=1$
given by $\rho(\lambda)=\int_{\mathbb{R}^d} \frac{d{\bf k}}{(2\pi)^d}\delta\left(\lambda-\mu-t{\bf k}^2\right)\propto t^{-1}((\lambda-\mu)/t)^{\frac{d-2}{2}} \theta(\lambda-\mu)$.}:
\begin{equation} \label{1dmatrix Anderson}
{\cal K}=\left(\begin{array}{ccccc} -t\frac{d^2}{dx^2}+\mu+W_{1,1}(x)&W_{1,2}(x)&\ldots & W_{1,N}(x)
\\ W_{1,2}(x)& -t\frac{d^2}{dx^2}+\mu+W_{2,2}(x)&\ldots & W_{2,N}(x)
\\ \ldots & \ldots &\ldots & \ldots
\\ \ldots & \ldots &\ldots & \ldots
\\  W_{1,N}(x) & \ldots & W_{N,N-1}(x) &  -t\frac{d^2}{dx^2}+\mu+W_{N,N}(x)
 \end{array} \right)
\end{equation}
Models of such type are sometimes called the matrix Anderson models, and are essentially continuous versions of Wegner orbital models.
Again, we will show below that the associated 'deformed semicircle' equation for the mean eigenvalue density of such problem can be
 solved  as long as $L\to \infty$ and yields an explicit form of the density profile.

\section{Summary of the main results}

In this paper our main object of interest is the disorder-averaged resolvent (Green's function) of the Hessian,
calculated at the absolute minimum ${\bf u}_0$ of the total energy:
\be \label{gf1}
\overline{{\cal G}(x,y;\lambda,{\bf u}_0)} = \frac{1}{N} \sum_{i=1}^N
\overline{ \left(\frac{1}{\lambda-K({\bf u}_0)} \right)_{xi,yi} }
\ee
as well as its limit at coinciding points $\overline{ {\cal G}(x,x;\lambda,{\bf u}_0)}$, which relates to the mean spectral density of the Hessian as
\be \label{dos}
\rho(\lambda)= \frac{1}{\pi} \lim_{{\rm Im}\,\lambda\to 0^-}\, {\rm Im} \,
\overline{{\cal G}(\lambda,{\bf u}_0)}
\quad , \quad  {\cal G}(\lambda,{\bf u}_0)= \frac{1}{N L^d} \sum_{x} \overline{ {\cal G}(x,x;\lambda,{\bf u}_0) }
\ee

Employing the replica trick, we first show that for $N \to \infty$ (the limit being taken for a fixed value of $L^d$) the average Green's function is given by
\be\label{Greendef}
\overline{ G(x,y;\lambda,{\bf u}_0)} = \int_k \frac{e^{i k (x-y)}}{\lambda -\mu_{\rm eff} + t \Delta(k) - 4 i p B''(0)}
\ee
where the value of the parameter $p$ is determined by the following self-consistent 'deformed semicircle' equation for the diagonal part
\be \label{Greendiag}
\overline{G(x,x;\lambda,{\bf u}_0)} = i p = \int_k \frac{1}{\lambda -\mu_{\rm eff} + t \Delta(k) - 4 i p B''(0)}
\ee
which is essentially of the same form as one for the orbital model with lattice Laplacian \cite{KhorPast1993}.

The only quantity which contains all the information about the optimization leading
to the ground state ${\bf u}_0$ is the parameter $\mu_{\rm eff}$.
 Below $\mu_{\rm eff}$ will be calculated in the various cases (replica-symmetric, 1RSB and FRSB) in the framework of the replica theory. We recall that the notation $\int_k$
applies both to the discrete models $\int_k = \frac{1}{L^d} \sum_k$
and the continuum limit $\int_k = \int \frac{d^d k}{(2 \pi)^d}$. These equations are quite general and
apply to basically arbitrary graph Laplacian matrices $t \Delta_{ix,jy}$ (even not translationally invariant
ones, provided the formula are
are generalized by replacing $\int_k \frac{1}{A} \to tr{A^{-1}}$, i.e.
the trace in internal space of the inverse matrix).

\subsection{Spectral Density of the Hessian at a generic point}

  As has been already mentioned, with setting $\mu_{eff}=\mu$ the above expressions (\ref{Greendef}, \ref{Greendiag})  provides the mean resolvent and the mean  spectral density $\rho(\lambda)$ for the manifold Hessian around a generic point of the disordered landscape.
Such object is interesting by itself and we study the shapes of the spectral density in detail for several examples.
Its generic feature is the square-root singularity at the spectral edges, which is thus a universal characteristics of the mean-field type spectral densities for disordered elastic systems of any dimension $d$. The shape as a whole is not universal and essentially depends on the dimension and the type of the Laplacian matrix (discrete or continuous).

\begin{figure}
\centering
\includegraphics[scale=.8]{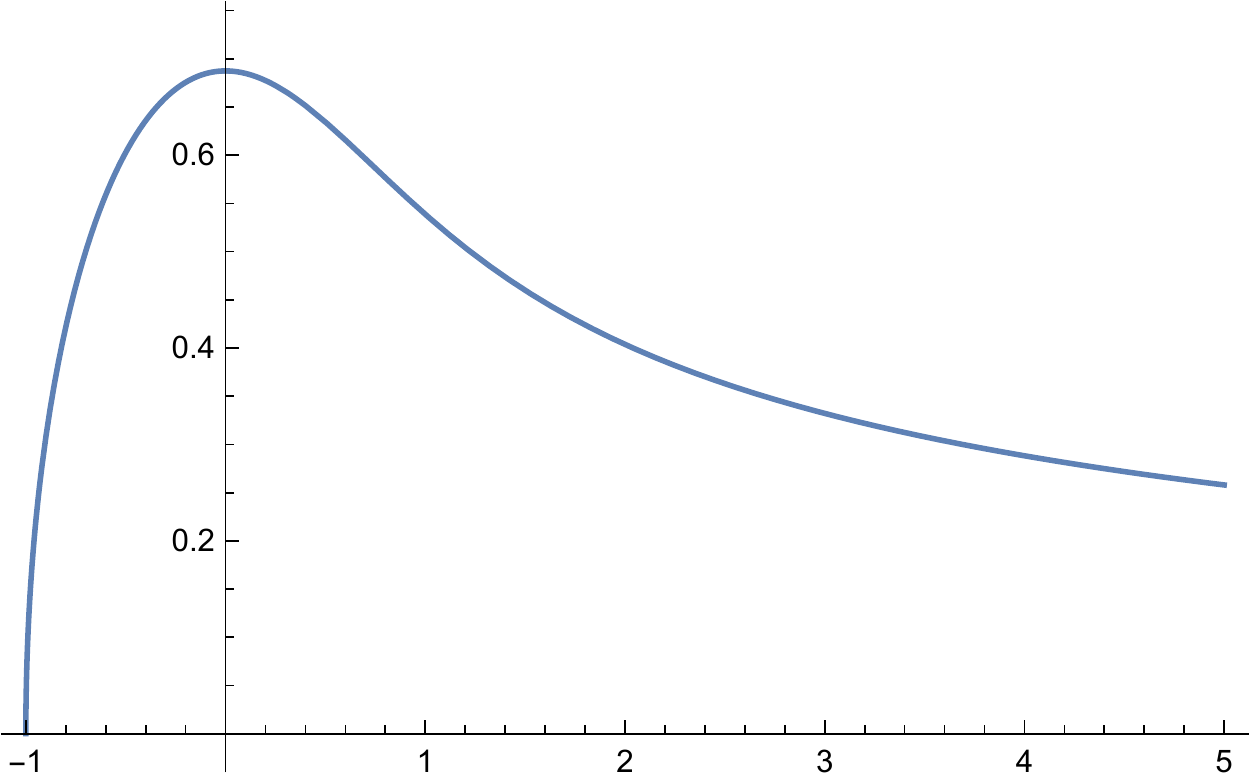}
\caption{Scaling function for the Hessian spectral density for the $d=1$ continuum model, $r_c(\Lambda)$
plotted versus $\Lambda=t^{1/3} \frac{\lambda-\mu_{\rm eff}}{3 B''(0)^{2/3}}$, as defined in \eqref{den1dcontexplicitintro}.}
\label{fig3}
\end{figure}

As relatively few explicit formulas are available in the literature for
eigenvalue densities of disordered matrices and operators beyond Wigner semicircular,  Marchenko-Pastur and 1D chains ( see
the book \cite{LifGrePas88} for those and further examples) we want to emphasize
that in our model
 it turns out to be possible to find explicitly the spectral density for the 1D matrix Anderson model  (\ref{1dmatrix Anderson}), of infinite length $L\to \infty$ and the Laplacian spectrum $-\Delta(k)=k^2, \, -\infty<k<\infty$:
\be\label{den1dcontexplicitintro}
\rho(\lambda) = \frac{1}{2\pi (t\,B''(0))^{1/3}} r_c\left(\Lambda=t^{1/3} \frac{\lambda-\mu}{3 B''(0)^{2/3}}\right), \quad r_c(\Lambda)=\frac{w_r^2}{4}\sqrt{\left(\frac{2}{w_r}\right)^3-1}
\ee
where
\be\label{card1dcontintro}
w_r=\left[1+\sqrt{1+\Lambda^3}\right]^{1/3}+\left[1-\sqrt{1+\Lambda^3}\right]^{1/3},
\ee
We have plotted in the Fig. \ref{fig3} the parameter free scaling function $r_c(\Lambda)$.
The spectral edge $\Lambda_e$ is given in this case by  $\Lambda_e=-1$. The function $r_c(\Lambda)$
reaches its maximum at $\Lambda=0$ and then decays at $\Lambda\gg 1$ as  $r_c(\Lambda\gg 1) \sim \frac{1}{\sqrt{3 \Lambda}}$.
The latter regime corresponds to the spectral density $\rho(\lambda)=\frac{1}{2\pi}\frac{1}{\sqrt{t(\lambda-\mu)}}$ of the disorder-free operator $\mu-\frac{d^2}{dx^2}$ with the spectrum $\lambda=\mu+tk^2$.

\begin{figure}
\centering
\includegraphics[scale=.8]{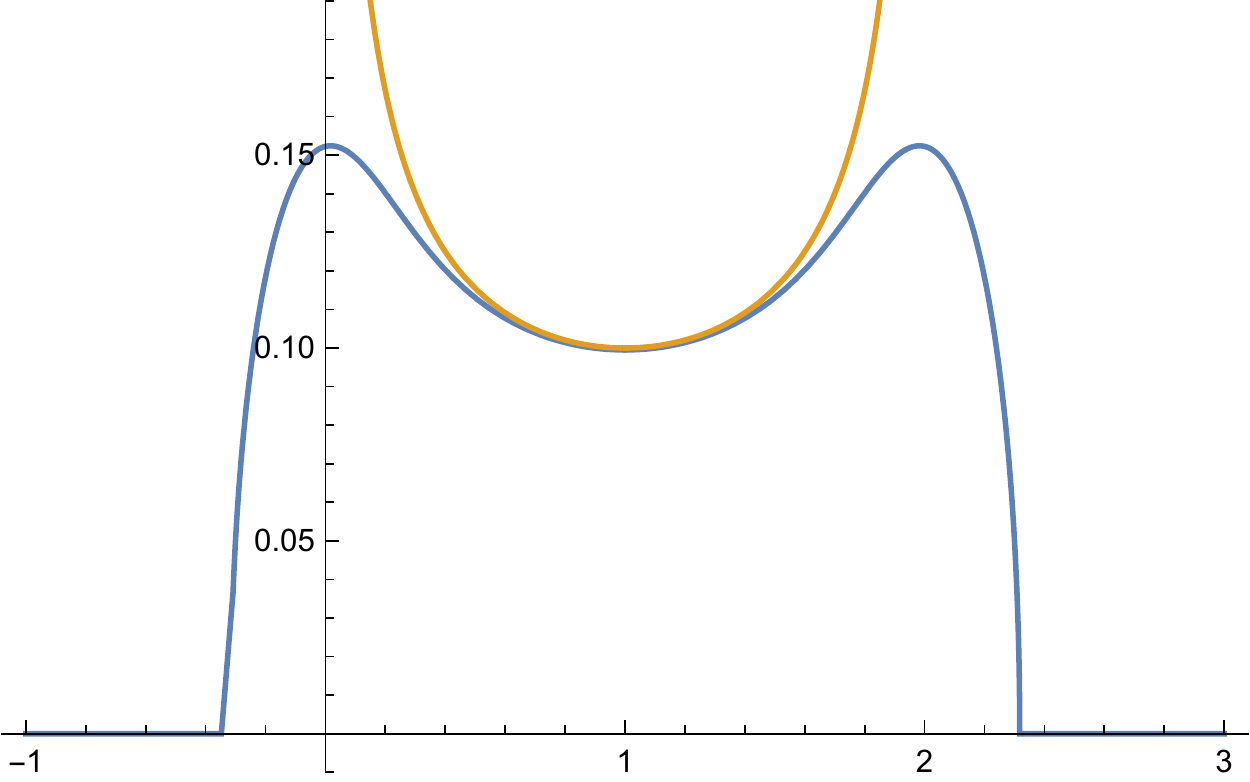}
\caption{Blue: scaling function for the Hessian spectral density, $r(\Lambda,y)$ versus
$\Lambda=\frac{\lambda- \mu }{2 t}$  for the infinite discrete 1D chain given
by Eq. \eqref{dens1intro},
for $y=\frac{t^2}{B''(0)} = 10$ (weak disorder).  In the weak disorder limit, the
central part converges to the spectral density without disorder (indicated here in orange),
while the two parts around the edges converge, upon rescaling, to the density for the continuum
model plotted in Fig. \ref{fig3}, according to Eq. \eqref{converge}.}
\label{fig2}
\end{figure}

In the case of 1D disordered elastic discrete chain with $-\Delta(k)=2(1-\cos{k}), 0\le k\le 2\pi$
the shape of the spectral density for the associated banded Hessian (and hence for the related Wegner orbital model) can be shown to be of the form
\be \label{dens1intro}
\rho(\lambda) =  \frac{t}{2 \pi B''(0)} r\left(\Lambda=\frac{\lambda- \mu}{2 t},y\right) \, ,y = \frac{t^2}{B''(0)}
\ee
but the function $r(\Lambda,y)$ does not have a simple form for $y\sim 1$. However, in the limiting case of weak disorder $y \gg 1$
a very explicit characterisation is again possible. In this case the graph $r(\Lambda,y)$ has two spectral edges at $\Lambda_e^{(-)}=-\frac{3}{2}y^{-2/3}$ and $\Lambda_e^{(+)}=2+\frac{3}{2}y^{-2/3}$
and the density profile in the vicinity of the edges is simply related
to the density profile $r_c(\Lambda)$ of $1D$ continuous system. Namely, in the vicinity of the left edge
\be \label{converge}
r(\Lambda,y)\approx y^{-2/3} ~ r_c\left(\frac{2}{3}y^{2/3}\Lambda\right), \quad |\Lambda|\sim y^{-2/3}
\ee
and essentially the same profile in the vicinity of the upper edge $|\Lambda-2|\sim y^{-2/3}$.
In between the edges, for any finite $0<\Lambda<2$ the profile for $y>>1$ is given by the ''disorder-free'' shape
$r(\Lambda,y) \simeq 1/(y \sqrt{\Lambda (2-\Lambda)})$. The numerically calculated spectral density for $y=10$ is presented
in Fig. \ref{fig2} and shows all those features.

After this digression about the Hessian spectral densities at a generic point of the disordered landscape, we return to our main task
of analysing the Hessian spectra conditioned by the requirement of sampling at the global minimum in the landscape, which requires the determination of $\mu_{eff}$. Before briefly summarizing our main results, we need to be more specific about the
correlations of the landscape, i.e. the choice of $B(q)$ in \eqref{cov}. The corresponding discussion is given below.

\subsection{Correlations of the random landscape and main features of the phase diagram}

For a general classification of the functions $B(q)$ corresponding to allowed covariances of
isotropic stationary Gaussian fields we refer to \cite{UsToy} and references therein. Here, for
applications to elastic manifolds we mainly consider the power-law class when the derivative $B'(q)$ can be written
as in \cite{LDMW}
\footnote{We follow the definitions and notations of \cite{LDMW}
(section II A and B) which are consistent with the original paper \cite{MP2}. The parameter
$\gamma$ is thus identical to the one defined in our previous work \cite{UsToy}.}:
\bea \label{defgamma}
B'(q) = - \frac{B_0}{ r_f^2 (1+ \frac{q}{\gamma r_f^2})^{\gamma}}, \quad \gamma>0
\eea
As special limiting cases this class also includes the (i) exponential $B(z)=B_0 e^{-z/r_f^2}$ as the limit $\gamma \to +\infty$,
and (ii) the log-correlated case for $\gamma\to 1$. Here $r_f$ is the correlation length
of the random potential which enters in Larkin's theory, and $B_0$ has dimension of energy square.
For notational simplicity we will consider
\bea
B(q) = A (c + q)^{\gamma-1}
\eea
hence choosing $c=\gamma r_f^2$ and $A=B_0 c^\gamma/((\gamma-1) r_f^2)$.\\

Let us recall the main features of the replica solution \cite{MP2,LDMW} for $N \to \infty$
(restricting for simplicity to $d \leq 4$).
Let us first define the ``Flory'' roughness and the free energy fluctuation exponents
\be
\zeta = \zeta_F(\gamma)=\frac{4-d}{2 + 2 \gamma} \quad , \quad
\theta= \theta_F(\gamma) = d-2 + 2 \zeta_F(\gamma)
\ee
Then it was found that for $\mu<\mu_c(T)$, full replica symmetry breaking, FRSB, occurs whenever
$\theta_F(\gamma) >0$, and 1-step replica symmetry breaking, 1RSB, occurs
when $\theta_F(\gamma) \leq 0$. The first case, FRSB, thus always occurs for manifold of dimensions $2 < d<4$, whereas
for $0\le d<2$ it is possible whenever $0<\gamma < \gamma_c(d)=\frac{2}{2-d}$. In that
case the exponents $\zeta,\theta$ (which are defined in the limit $\mu \to 0$) are
given by their Flory values. In the limit $\mu \to 0$ the system was shown to remain
in the glass FRSB phase at any temperature $T$ (no transition).
The second case, 1RSB, occurs for $d<2$ and $\gamma>\gamma_c(d)$. In that
case there is a phase transition at $T_c(\mu)$ which survives for
$\mu=0$. It is worth mentioning that in the marginal case $\gamma=\gamma_c(d)$
this transition is of a continuous nature.

The exponents are $\theta=0$ and $\zeta=\frac{2-d}{2}$
in both the high-T phase and the low-T 1RSB phase, with however
different amplitudes
\footnote{This is an artefact of $N=\infty$ for 1RSB and may not
survive for $N=1$, except maybe in the boundary case $\gamma=\gamma_c(d)$
(certainly it survives for $d=0$, $\gamma_c=1$ the log-correlated case).}.
The special case $\gamma = \gamma_c(d)$
is called marginal and exhibits features of both 1RSB and FRSB. Note that it also includes as a special limit the case of $d=2$ and the disorder with exponential covariance.

In \cite{UsToy}, for the case of a single particle $d=0$, we have distinguished
long-range correlated (Full RSB) $0<\gamma<1$,
and short-range correlated (1-RSB) $\gamma>1$ landscapes.
For the manifold such as distinction thus also holds, however
the critical value of $\gamma$ is not unity anymore, but
equal to $\gamma_c(d)=\frac{2}{2-d}$. In particular
for $d>2$ one is always in the LRC case
\footnote{Note however that for $d>4$ there is again a
RS phase for weak disorder.}. This is because
the total energy now also includes the elastic energy,
which increases the correlations of the
{\it effective} random landscape seen by the manifold.

\subsection{Hessian spectrum at the point of global energy minimum}

Our results here extend the ones of \cite{UsToy}, which are recovered in the special case
of $d=0$. There are many similarities with that case. The most important parameter
in the theory is the "Larkin mass" $\mu_c>0$ which controls the value of the parabolic confinement $\mu$ below which
 the replica symmetry breaking (RSB) occurs. Its value turns out to be given by the {\it positive} solution
 of
\be\label{LMintro}
1 = 4  B''(0) \int_k \frac{1}{(-  t \Delta(k) + \mu_c)^2}
\ee
which is controlled both by disorder strength and the elasticity matrix.
For example, for $1D$ continuous system a simple calculation gives $\mu_c= \left(\frac{B''(0)}{\sqrt{t}}\right)^{2/3}$.
Our analysis shows that in the replica symmetric phase
the {\it lower} spectral edge $\lambda^{(-)}_e$  of the Hessian (which we associate with the spectral gap)
as a function of $\mu$ is given by
\bea \label{edgelowintro}
\lambda^{(-)}_e =  \mu - \mu_c
+ 4 B''(0)  \int_k \left[ \frac{1}{\mu - t \Delta(k)}  - \int_k \frac{1}{\mu_c - t \Delta(k)} \right]
\eea
This formula immediately shows that for $\mu>\mu_c$ the Hessian spectrum is always gapped (from zero).
Upon expanding for $\mu \to \mu_c$ and using (\ref{LMintro}) one immediately finds the gap vanishing {\it quadratically} at $\mu_c$.
For $\mu<\mu_c$ the Hessian spectrum is either gapped or extends down to zero,
depending if 1-step RSB or full-RSB occurs. In the first case, the gap vanishes
as $(\mu_c-\mu)^4$ near the transition from below, with the super-universal exponent. For example,
 for the continuum model in dimension $d$ we get for $\mu=\mu_c(1-\delta)$
\be\label{1RSBgapintro}
\lambda_e^{(-)} = \frac{\mu_c}{36 B^{(3)}(0)^4} \left(\frac{4-d}{4}\right)^3
\left(B^{(4)}(0) B''(0) - \frac{2 (3-d)}{4-d} B^{(3)}(0)^2
\right)^2  \delta ^4 +O\left(\delta ^5\right)
\ee
In the second case of full RSB the gap is identically zero everywhere for $\mu\le\mu_c$.

We also obtain the average two point Green function \eqref{gf1}
and we find that at the edge of the spectrum it decays exponentially as $\sim e^{-|x-y|/L_c}$,
with the characteristic length precisely equal in all cases to the Larkin length $L_c$
introduced in the theory of pinning. For the continuum model with short-range elasticity and weak disorder,
$L_c \sim 1/\mu_c^{1/2}$. This is thus reminiscent of the results of
\cite{CaoRossoSoftModes} although obtained there in a slightly different context
(depinning). Remarkably, this property holds also for
$\mu>\mu_c$, i.e. in the RS phase.

As a by product of these studies we arrived to a very precise criterion which allows to
determine which types of covariance functions $B(q)$ in a given manifold dimension $d$ will lead
 to the full-RSB solution. It reads
\be\label{criterionintro}
A(q) = \frac{2(3-d)}{4-d} \left(B'''(q)\right)^2 - B''(q) B''''(q) < 0 \quad \Leftrightarrow \quad \text{Full RSB}
\ee
which generalizes the criterion given in \cite{FS2007} for $d=0$.
Inserting $B(q)$ for the power law models \eqref{defgamma} gives a criterion in agreement
with one given in \cite{MP2}, namely that the full RSB solution holds (i) for any value of $\gamma$ if $d \geq 2$ and
(ii) for $\gamma \leq \gamma_c(d)=2/(2-d)$ if $d \leq 2$.

Finally, for $d=1$ the above criterion specifies the covariance $B(q) = \frac{A}{c + q}$ as the special marginal case which shares simultaneously the features of 1RSB and FRSB, and  for $d=2$
 the exponential $B(q) \sim e^{-a q}$ plays a similar role. In particular, the Hessian spectrum is gapless in those potentials.
  We study both cases in much detail and show that for them the Parisi equations can be solved exactly and explicitly. Note that in $N\to \infty$ class of models these cases play the same role for $d=1$ and $d=2$ as the logarithmically correlated case identified as marginal in $d=0$ \cite{FS2007}. It is worth mentioning here that due to marginality many special properties of logarithmically correlated potential in $d=0$ survive for finite $N$, as was originally suggested in \cite{CLD}
and much studied in the last decade, see e.g. \cite{FB2008,FLDR,FyoKeat2014}. It would be
interesting to investigate whether some universality holds for the finite-$N$ elastic disordered systems in the above marginally correlated cases for $d=1,2$ as well.\\

 Let us mention here some  works on related  models,
although they are more similar to the case $d=0$, and not the
manifold. In \cite{RBBC2018,YamVil2018} the Hessian statistics is sampled over all saddle-points or minima at a given value of the potential ${\cal H}({\bf u})=E=const$, a priori quite different
from imposing the absolute minimum. The spectrum of the soft modes was also calculated in a mean-field model of the jamming transition, the 'soft spherical perceptron'. The Hessian matrix in that model has the shape of a (uniformly shifted) Wishart matrix, whose spectrum is given by the
(shifted) Marchenko-Pastur  law, while in \cite{UsToy} the Hessian spectrum is given by a shifted Wigner semicircle.
The model has two phases: ' RS simple' and 'FRSB complex' and the Marchenko-Pastur spectrum  in that model was demonstrated to undergo a transition from gapped to gapless, similar to what we find here for Gaussian landscapes. Finally, it is worth mentioning a quite detailed recent characterization of the energy landscape of spherical $p-$spinglass in full-RSB phase close to the global minimum, see  \cite{Subag18ground} and references therein.

The outline of this paper is as follows.
In Section \ref{sec:derGreen} we provide a derivation of the average
Green function, resolvent and the spectral density of the Hessian using two sets of replica.
The second set is necessary to specify that the Hessian is considered at the absolute energy minimum.
We obtain the general saddle point equations which determine these quantities.
In Section \ref{sec:analysis} we analyze the results. In the first subsection \ref{sec:dos}
we obtain the spectral density and the Green function keeping $\mu_{\rm eff}$ as a free
parameter. The general results only weakly depend on this parameter, which simply
globally shifts the support of the spectral density. In the second part \ref{sec:phases} we complete
the study by calculating $\mu_{\rm eff}$ from the explicit solution of the replica
saddle point equations. This leads to the determination of the spectral edges and of the gap, in the three main distinct cases: replica symmetric, FRSB and 1RSB. The case of marginal 1RSB is given a special attention. Finally
Section \ref{sec:conclusion} contains the conclusion.

\section{Derivation of the average Green function using replica}
\label{sec:derGreen}

Below we use the following notational conventions. The sums over the internal points of the manifold $x,y,..$ are denoted as $\sum_x \equiv \sum_{x=1}^{L^d}$,
the sum over the first set of replica indices $\alpha,\gamma,..$ are denoted
$\sum_\alpha \equiv  \sum_{\alpha=1}^m$, the sums over the second set of replica indices
$a,b,c..$ are denoted $\sum_a \equiv  \sum_{a=1}^m$, and similarly for the products.
The indices $i=1,..N$ and the dot product is used in $\mathbb{R}^N$.

The notation $\Tr$ is the trace over all indices $x,i$ and $a$ or $\alpha$, i.e.
over $\mathbb{R}^{m} \times L^d$ or $\mathbb{R}^{n} \times L^d$,
e.g. $\Tr A = \sum_{xa} A_{xa,xa}$. The notation $\tr$ is reserved for the traces over $a$ or $\alpha$ only,
i.e. over $\mathbb{R}^{m}$ or $\mathbb{R}^{n}$, i.e. $\tr A= \sum_{a} A_{aa}$.

\subsection{Green's function and the first set of replica}

As the starting point of our approach, we introduce the
resolvent of the Hessian  $K({\bf u})$ defined in \eqref{Hessian}, for a given generic configuration
${\bf u}(x)$ (not necessarily the minimum of the total energy) and in a given
realization of the random potential  $V(\mathbf{u}(x),x)$.  The associated Green's function is then defined via
\be
{\cal G}(x,y;\lambda,{\bf u})= \frac{1}{N} \sum_{i=1}^N \left(\frac{1}{\lambda-K({\bf u})} \right)_{xi,yi}
\quad , \quad {\cal G}(\lambda,{\bf u})= \frac{1}{N L^d} \sum_{x} {\cal G}(x,x;\lambda,{\bf u})
\ee
 Such Green's function
admits then the following representation in terms of $m$ replicated Gaussian integrals over $N-$component real-valued vector fields $\phi_{\alpha}(x)$, with $\alpha=1,\ldots,m$:
\be\label{1streplicanew}
{\cal G}(x,y;\lambda,{\bf u})= \lim_{m \to 0} \int_{\mathbb{R}^{Nm}}
e^{-\frac{i}{2} \lambda \sum_{x,\alpha} \phi^2_\alpha(x)}
e^{ \frac{i}{2} \sum_{x,y,\alpha} \phi_\alpha(x) \cdot K({\bf u}) \cdot \phi_\alpha(y)  }\,
\ee
\[
\times \left[\frac{i}{m N} \sum_{\gamma} \phi_{\gamma}(x) \cdot \phi_{\gamma}(y)\right]\,\prod_{x,\alpha} {\cal D} \phi_{\alpha}(x)
\]
where we assumed that ${\rm Im}\,\lambda<0$ and
set the factor $(\frac{i}{\pi})^{m/2} \to 1$ for $m=0$. From this we calculate the mean spectral density of the Hessian eigenvalues ``at a temperature $T$'', defined as
\be \label{rhoT}
\rho_T(\lambda)=\frac{1}{\pi}  \lim_{{\rm Im}\,\lambda\to 0^-}\, {\rm Im} \,
\overline{\left\langle {\cal G}(\lambda,{\bf u})\right\rangle_T}
 \ee
where the thermal averaged value of any functional $g({\bf u})$ of a configuration ${\bf u}(x)$  at a temperature $T=\beta^{-1}$ is defined
as $\langle g({\bf u}) \rangle_T :=\int g({\bf u}) \pi_{\beta}({\bf u}){\cal D}{\bf u}(x)$, with
$\pi_{\beta}({\bf u})={\cal Z}_{\beta}^{-1} e^{-\beta {\cal H}[{\bf u}]}$ being the Boltzmann-Gibbs weights
 associated with the configurations via the energy functional (\ref{landscape}). Our final aim is then to obtain the mean spectral density of Hessian eigenvalues at the
 absolute minimum by setting temperature to zero:
 \be
\rho(\lambda)= \lim_{T \to 0} \rho_T(\lambda) = \frac{1}{\pi} \lim_{{\rm Im}\,\lambda\to 0^-}\, {\rm Im} \,
\overline{{\cal G}(\lambda,{\bf u}_m)}
 \ee
The problem therefore amounts to first calculating the disorder and thermal average
\be\label{1streplicanew1}
\overline{ \left\langle {\cal G}(x,y;\lambda,{\bf u}) \right\rangle_T} =  \lim_{m \to 0} \int_{\mathbb{R}^{Nm}}
e^{-\frac{i}{2} \lambda \sum_{x,\alpha} \phi^2_\alpha(x)}\,
\overline{ \langle e^{ \frac{i}{2} \sum_{x,y,\alpha} \phi_\alpha(x) \cdot K({\bf u}) \cdot \phi_\alpha(y) }\rangle_T}
\ee
\[
\times \left[\frac{i}{m N} \sum_{\gamma} \phi_\gamma(x) \cdot \phi_\gamma(y)\right]\,\prod_{x,\alpha} {\cal D}\phi_\alpha(x)
\]
where
\begin{equation}\label{thermavenew}
\langle e^{ \frac{i}{2} \sum_{x,y,\alpha} \phi_\alpha(x) \cdot K({\bf u}) \cdot \phi_\alpha(y)  }  \rangle_T =
{\cal Z}_{\beta}^{-1} \int_{\mathbb{R}^N}  \, {\cal D}{\bf u}(x) \,\, e^{ \frac{i}{2} \sum_{x,y,\alpha} \phi_\alpha(x) \cdot K({\bf u}) \cdot \phi_\alpha(y) -\beta {\cal H}[{\bf u}]}
\end{equation}
and then, by performing the zero-temperature limit, to capture the contribution from the global minimum configuration only.\\

\subsection{Average Green function and second set of replica}

In the framework of the replica trick we represent the
normalization factor ${\cal Z}^{-1}_{\beta}$ in Eq.(\ref{thermavenew})
formally as $1/{\cal Z}_{\beta}=\lim_{n\to 0}{\cal Z}_{\beta}^{n-1}$
and treat the parameter $n$ before the limit as a positive integer. After this is done, averaging the product of
$n$ integrals over the Gaussian potential $V({\bf u})$ is an easy task. The calculation is very similar
to the one for $d=0$ in \cite{UsToy}, (apart from an additional factor of $2$ for each derivative of
$B$ arising due to a slightly different normalization of the covariance used in \cite{UsToy} ) and we simply quote the result
referring to \cite{UsToy} for more details. We obtain
\be \label{form1new}
\overline{ \langle e^{ \frac{i}{2} \sum_{x,y,\alpha} \phi_\alpha(x) \cdot K({\bf u}) \cdot \phi_\alpha(y)  }  \rangle_T }
= \lim_{n \to 0}  \int \, \prod_{x,a} {\cal D}{\bf u}_a(x) \, e^{- L_{n,m}[\bf u, \phi] }
\ee
with
\begin{equation}\label{actionnew}
L_{n,m}[{\bf u}, \phi]={\cal L}_{m}[\phi]+{\cal L}_{n,m}[{\bf u}, \phi]
\end{equation} where the
${\bf u}$-independent part of the action is given by
\bea \label{thermave1Anew}
\!\!\!\!\!\!\!\!\!  {\cal L}_{m}[\phi] &=& \sum_x \frac{B''(0)}{2N}\left[ \left(\sum_{\alpha} \phi_\alpha^2(x)\right)^2
+2 \sum_{\alpha,\beta} \left(\phi_\alpha(x) \cdot \phi_\beta(x)\right)^2 \right]
\\
\!\!\!\!\!\!\!\!\!  &-& \frac{i}{2} \sum_{x,y,\alpha} \phi_\alpha(x) \cdot (\mu \mathbf{1} - t \Delta)_{xy} \phi_\alpha(y)
\nonumber
\eea
whereas both
${\bf u}$- and $\phi-$ dependent part is
\bea \label{thermave1A2new}
\fl {\cal L}_{n,m}[{\bf u},\phi]&=&\frac{\beta}{2}\sum_{x,y,a} {\bf u}_a(x)
\cdot (\mu \mathbf{1} - t \Delta)_{xy} {\bf u}_b(y)
-N\frac{\beta^2}{2}\sum_{x,a,b} B\left(\frac{\left({\bf u}_a(x)-{\bf u}_b(x)\right)^2}{N}\right)
\nonumber \\
\fl &
 +& i\beta \sum_x (\sum_\alpha \phi_\alpha(x)^2)\,\sum_{a} B'\left(\frac{\left({\bf u}_a(x)-{\bf u}_1(x)\right)^2}{N}\right)
\\
\fl &
+& \frac{2 i\beta}{N}\sum_x \sum_{a} B''\left(\frac{({\bf u}_a(x)-{\bf u}_1(x))^2}{N}\right)
\sum_{\alpha}
\left( \left({\bf u}_a(x) -{\bf u}_1(x) \right) \cdot \phi_\alpha(x) \right)^2 \nonumber
\eea
We can thus rewrite
\be\label{1streplicanew2}
\overline{ \left\langle {\cal G}(x,y;\lambda,{\bf u}) \right\rangle_T} = \lim_{m,n \to 0} \int_{\mathbb{R}^{Nm}}
\prod_{x,\alpha} {\cal D}\phi_\alpha(x)  \int \, \prod_{x,a} {\cal D}{\bf u}_a(x)\,
e^{- \frac{i}{2} \lambda \sum_{x,\alpha} \phi^2_\alpha(x) - L_{n,m}[\bf u, \phi] }
\ee
\[
\times \frac{i}{m N} \left[\sum_{\gamma} \phi_\gamma(x) \cdot \phi_\gamma(y)\right]
\]

Now we introduce auxiliary fields and their conjugate fields. We define for each value of the argument $x$ (which we omit for brevity) the differentials
\bea
\fl && ~~~~ dQ=\prod_{1 \leq a \leq b \leq n} dQ_{ab} \quad , \quad
dP=\prod_{1 \leq \alpha \leq \beta \leq m} dP_{\alpha \beta}
\quad , \quad
dR=\prod_{a=1}^n \prod_{\alpha=1}^m dR_{a \alpha} \\
\fl && ~~~~ d\sigma=\prod_{1 \leq a \leq b \leq n} d\sigma_{ab} \quad , \quad
d\tau=\prod_{1 \leq \alpha \leq \beta \leq m} d\tau_{\alpha \beta}
\quad , \quad
d\eta=\prod_{a=1}^n \prod_{\alpha=1}^m d\eta_{a \alpha}
\eea
and then define ${\cal D}Q(x)=\prod_x\,dQ(x)$, etc. This allows us to use the formal identity
\bea
\fl  1 &=& \int \prod_x {\cal D}Q(x){\cal D} \sigma(x) {\cal D}P(x) {\cal D}\tau(x) {\cal D}R(x) {\cal D}\eta(x) \,
e^{- \frac{\beta}{2} \sum_{x,a,b} \sigma_{ab}(x)\left(N Q_{ab}(x) - {\bf u}_a(x) \cdot {\bf u}_b(x) \right)}
\nonumber \\
\fl &
\times& e^{\frac{i}{2} \sum_{x,\alpha,\beta} \tau_{\alpha \beta}(x) \left(N P_{\alpha \beta}(x)- \phi_\alpha(x) \cdot \phi_\beta(x)\right)
+ \frac{1}{2} \sum_{x,a,\alpha} \eta_{\alpha \beta}(x) \left(N R_{\alpha \beta}(x)- {\bf u}_a(x)  \cdot \phi_\alpha(x)\right)
}
\eea
where the contours of integration are duly chosen. This identity can be then inserted inside \eqref{1streplicanew2} effectively allowing to replace all scalar products
${\bf u}_a(x) \cdot {\bf u}_b(x)$, $\phi_\alpha(x) \cdot \phi_\beta(x)$
and ${\bf u}_a(x)  \cdot \phi_\alpha(x))$ by $Q_{ab}$, $P_{\alpha \beta}$ and
$R_{a,\alpha}$ respectively inside ${\cal L}_{n,m}[{\bf u},\phi]$, leaving
a simple quadratic form in the
fields ${\bf u}_a(x)$ and $\phi_\alpha(x)$, which can be integrated out.
Restricting for now, for simplicity, to the diagonal element of the resolvent
we obtain
\bea
\fl && \overline{ \left\langle {\cal G}(x,x;\lambda,{\bf u}) \right\rangle_T} \\
\fl && = \lim_{m,n \to 0}
\int \prod_x {\cal D}Q(x) {\cal D}\sigma(x) {\cal D}P(x) {\cal D}\tau(x) {\cal D}R(x) {\cal D}\eta(x)  \left[\frac{i \tr P(x)}{m N}\right]  e^{- N L[Q,\sigma,P,\tau,R,\eta] } \nonumber
\eea
where we have defined the action
\be
L[Q,\sigma,P,\tau,R,\eta]= L_m[P,\tau] + L_{n,m}[Q,\sigma,P] + \delta L[Q,\sigma,P,\tau,R,\eta]
\ee
\be\label{actPnew}
L_m[P,\tau]= \sum_x \frac{B''(0)}{2}\left[\left(\tr\,P(x)\right)^2+2\tr\left(P(x)^2\right)\right]+\frac{i \lambda}{2} \tr P(x)
\ee
\[+ \frac{1}{2} \Tr \ln \left((\mu \mathbf{1} - t \Delta) \mathbf{1}_m - \tau \mathbf{1} \right)
- \frac{i}{2} \sum_{x} \tr (\tau(x) P(x))
\]
\[
L_{n,m}[Q,\sigma,P]=
 \sum_x \frac{\beta\mu}{2}\tr Q(x) -\frac{\beta^2}{2}
\sum_{a,b=1}^n B\left(Q_{aa}(x)+Q_{bb}(x)-2Q_{ab}(x) \right)
\]
\be \label{actQnew}
+ \frac{1}{2} \Tr \ln ((\mu \mathbf{1} - t \Delta) \mathbf{1}_n - \sigma \mathbf{1} )
\ee
\[
 + \frac{\beta}{2} \sum_{x} \tr (\sigma(x) Q(x))
+ i\beta \sum_x \tr P(x) \,\sum_{a=1}^n B'\left(Q_{aa}(x)+Q_{a1}(x)-2Q_{11}(x)\right)
\]
The last piece $\delta L[Q,\sigma,P,\tau,R,\eta]$ is given in Appendix A.
Since it vanishes at the saddle point we do not need to give it here.

\medskip

We can now write the saddle point equations.
It is easy to check that the equations admit an $x$-invariant solution in all
variables:
i.e.  $Q_{ab}(x)=Q_{ab}$, $\sigma_{ab}(x)=\sigma_{ab}$,
$P_{\alpha \beta}(x)=P_{\alpha \beta}$, $\tau_{ab}(x)=\tau_{ab}$ at the saddle point. We will consider only this solution on physical grounds. Let us define again the quantity
\be \label{mueffdef}
\mu_{\rm eff} = \mu - 2 \beta \sum_{a=1}^n B'\left(Q_{aa}+Q_{11}-2Q_{a1}\right)
\ee

Taking first the functional derivatives w.r.t. $\tau_{\alpha \beta}(x)$ and $P_{\alpha \beta}(x)$ we arrive at the saddle-point equations
\bea
 - i P_{\alpha \beta} &=& ((\mu \mathbf{1} - t \Delta) \mathbf{1}_m - \tau \mathbf{1} )^{-1}_{\alpha x, \beta x} \\
 \tau_{\alpha \beta}&=& - 4 i B''(0) (P_{\alpha \beta} + \frac{1}{2} \delta_{\alpha \beta} \tr P)
+ (\lambda  + \mu - \mu_{\rm eff} ) \delta_{\alpha \beta}
\eea
Moreover, similarly to $d=0$ case treated in detail in \cite{UsToy}, it is easy to check that the  invariance of the action under rotating
matrices $P$ and $\tau$ in the replica space implies that the corresponding saddle point solutions must be actually proportional to the identity matrix: $P_{\alpha \beta}=p \delta_{\alpha \beta}$ and
$\tau_{\alpha \beta}= \tau \delta_{\alpha \beta}$.  In the limit $m \to 0$ one then finds that $\tau$ satisfies the  equation
\be \label{tauself}
\tau = - 4 i B''(0) p + \lambda  + \mu - \mu_{\rm eff}
\ee
which when substituted to the corresponding equation for $p$ yields the closed self-consistency equation for the latter:
\bea
i p &=& - ((\mu \mathbf{1} - t \Delta) \mathbf{1}_m - \tau \mathbf{1} )^{-1}_{xx} \nonumber
\\
\label{pself}
& = & - \int_k \frac{1}{\mu - t \Delta(k) - \tau} =  \int_k \frac{1}{\lambda -\mu_{\rm eff} + t \Delta(k) - 4 i p B''(0)}
\eea
This condition has exactly the form of the 'deformed semicircle' equation derived in \cite{KhorPast1993} for the block-banded
Wegner orbital model, assuming the random matrices $W^{(r)}, \, r=1,\ldots, L$ on the main diagonal to be of standard GOE type.
  To that end it is worth noting that in the action eq.(\ref{actPnew}) eventually responsible for fixing the shape of the self-consistency equation the difference between our choice for  $W^{(r)}$, see (\ref{covWdisc}), and the GOE appears only via the term $(\tr P)^2$ absent for GOE case. However for $m\to 0$ that term gives contribution of the order $m^2$, hence is negligible in comparison with the dominant contributions of order $m$. Hence, from the point of view of calculating the profile of the mean eigenvalue density the difference between our Hessians and the Wegner orbital model is immaterial.
 In particular, in the case of $d=0$ one recovers the self-consistency
condition found in \cite{UsToy}
\be
i p = \frac{1}{\lambda - \mu_{\rm eff} -  4 i p B''(0)}
\ee
whose solution is the genuine semicircular density as typical for $d=0$ random matrix  problems.
The analysis of the solution to the self-consistency equation (\ref{pself}) for higher $d$, and especially for $d=1$ will be provided in detail below.

Along similar lines  one can derive the average Green's function at two different points $x\ne y$.
 Starting from its definition and using a source term, it is not difficult to see that in the limit of large $N$ employing the
same saddle point solutions one arrives at the following representation:
\be \label{Green}
\overline{{\cal G}(x,y;\lambda,{\bf u}_0)} = \int_k \frac{e^{i k (x-y)}}{\lambda -\mu_{\rm eff} + t \Delta(k) - 4 i p B''(0)}
\ee
where $p$ is an apriori complex number determined by the (self-consistent) equation for the diagonal part.

Finally, using this saddle point, we see that the term which couples $P$ and $Q$ is proportional
to $m$ and hence as $m\to 0$ can be neglected in the saddle point equation for $Q$.
The resulting equations are identical to those obtained in \cite{MP2,PLDKWLargeNDetails,LDMW}
and we now briefly recall them here. Taking the functional derivatives w.r.t. $\sigma_{ab}(x)$
and $\tau_{\alpha \beta}(x)$ yields
\be \label{repsp1}
 \beta Q_{ab} = (((\mu \mathbf{1} - t \Delta) \mathbf{1}_n - \sigma \mathbf{1})^{-1}_{ax,bx}
= \int_k G_{ab}(k) \\
\ee
\be \label{repsp1a}
\sigma_{ab}= 2\beta ( \sum_c B'(\chi_{ac}) - B'(\chi_{ab}) ) \quad  \Leftrightarrow
\quad \sigma_{ab}= - 2 \beta B'(\chi_{ab})
\quad \text{and} \quad \sum_b \sigma_{ab}=0
\ee
where we define, as in \cite{MP2,PLDKWLargeNDetails,LDMW}
\bea  \label{repsp2}
G_{ab}(k) = \left((\mu  - t \Delta(k)) \mathbf{1}_n - \sigma \right)^{-1}_{ab} \\
\chi_{ab}= T \int_k (G_{aa}(k) + G_{bb}(k) - 2 G_{ab}(k))\label{repsp2chi}
\eea
We will recall briefly the analysis of these equations below, as needed.

\section{Analysis of the results}
\label{sec:analysis}

We now analyze the results from these saddle-point equations in two stages. First we
analyse the general form of the average Green function and the spectral density by simply
assuming that $\mu_{\rm eff}$ takes some value at $T=0$. That gives the shape of the
spectral density  $\rho(\lambda)$, up to a global shift of $\lambda$.
In a second part, we recall the analysis leading to the various phases
(RS, FRSB and 1RSB) and obtain from it the corresponding possible values of $\mu_{\rm eff}$
as a function of $\mu$, which allows to determine the location of the edge of the spectrum.

\subsection{The spectral density and Green's function}
\label{sec:dos}

\subsubsection{General formula, Larkin mass and lower edge}

We start with recalling the self-consistency equation for the diagonal part of the Green's function, the parameter $p$:

\be \label{diag2}
\overline{G(x,x;\lambda,{\bf u}_0)} = i p = \int_k \frac{1}{\lambda -\mu_{\rm eff} + t \Delta(k) - 4 i p B''(0)}
\ee
There are usually multiple solutions for $p$ and we must choose the branch such that for $\lambda \to \pm \infty$ one has $i p \sim \frac{1}{\lambda}$.
For a discrete model the spectrum of the perturbed Laplacian is bounded and  large $|\lambda|$ necessarily correspond
to being outside of the spectrum. In the continuum model the same holds for large negative $\lambda$.
In the range of $\lambda$ outside of the spectrum, $p$ is necessarily pure imaginary. When
$\lambda$ reaches the edges of the spectrum and goes inside the spectral support, $p$ develops a real part proportional
to the mean spectral density. Hence we can write for real $p_1,p_2$
\bea\label{denim}
p = p_1- i p_2, \quad \quad \rho(\lambda) = \frac{1}{\pi} {\rm Im}(i p) =
\frac{1}{\pi}  p_1
\eea
which converts \eqref{diag2} after separating the real and imaginary parts, into two coupled equations which determine $p_1$ and $p_2$ as functions of $\lambda$:

\be \label{systemeq1}
p_2 = \int_k \frac{\lambda -\mu_{\rm eff} + t \Delta(k) - 4 p_2 B''(0)}{
( \lambda -\mu_{\rm eff} + t \Delta(k) - 4 p_2 B''(0))^2 + \left[4p_1 B''(0)\right]^2}:=F(\lambda,p_2,p_1^2)  \\
\ee
\be p_1 = 4 p_1 B''(0) \int_k \frac{1}{
( \lambda -\mu_{\rm eff} + t \Delta(k) - 4 p_2 B''(0))^2 + \left[4p_1B''(0)\right]^2}:=p_1 G(\lambda,p_2,p_1^2)
\label{systemeq2}
\ee

The edge of the spectrum is
for $\lambda=\lambda_e$ such that $p_1$ acquire a non zero value, hence it is determined by
eliminating $p_2=p^e_2$ in the system
\bea \label{eliminate0}
&& p_2^e = \int_k \frac{1}{
 \lambda_e -\mu_{\rm eff} + t \Delta(k) - 4 p_2^e B''(0)}  \\
&&1 = 4  B''(0) \int_k \frac{1}{
( \lambda_e -\mu_{\rm eff} + t \Delta(k) - 4 p_2^e B''(0))^2}\label{eliminate00}
\eea
since at the edge one can set $p_1=p_1^e=0$. Note that there
can be more than one edge, i.e. more than one solution
to this system. Note also that assuming the right-hand-sides in (\ref{systemeq1}) and (\ref{systemeq2})  are analytic functions $F,G$ of all arguments $\lambda, p_2$ and $u=p_1^2$, a straightforward expansion in powers of $\lambda-\lambda_e$ shows that just above the lower edge
\be\label{edgebehaviour}
p_1^2\approx (\lambda-\lambda_e) \frac{ \frac{\partial G}{\partial p_2}|_e\frac{\partial F}{\partial \lambda}|_e-\left(\frac{\partial F}{\partial p_2}|_e-1\right) \frac{\partial G}{\partial \lambda}|_e }
{ (\frac{\partial F}{\partial p_2}|_e-1)\frac{\partial G}{\partial u}|_e -\frac{\partial F}{\partial u}|_e\frac{\partial G}{\partial p_2}|_e}
\ee
implying a square-root singularity of the density of eigenvalues at the thresholdin the generic case (where neither the numerator or denominator vanishes in \eqref{edgebehaviour}).

To further analyze these equations we introduce the {\it Larkin mass} $\mu_c>0$ defined as the {\it positive} solution
 of
\be\label{Lark}
1 = 4  B''(0) \int_k \frac{1}{(-  t \Delta(k) + \mu_c)^2}
\ee
Anticipating a little on the subsequent analysis, \eqref{Lark} precisely determines the range of curvatures when the replica-symmetric solution becomes unstable. Namely, it becomes unstable in the interval $0\le \mu<\mu_c$, with $\mu_c$ determined by
\eqref{Lark}. The Larkin mass exists whenever $B''(0) > 1/(4 \int_k \frac{1}{-  t \Delta(k)})$ and when
this is the case, it is unique. In the opposite case, the RS solution is stable for
all values of $\mu$. Note that $\mu_c$ depends only on $B''(0)$ and on the graph Laplacian
elasticity matrix.

We now assume that we are in the first case and there exists a finite Larkin mass $\mu_c>0$. It is then easy to find a solution
for the spectral edge $\lambda_e$. One sees that $p_2^e$ is now determined in terms of $\mu_c$ and that
\eqref{eliminate0} is equivalent to
\be  \label{simple}
\lambda_e := \lambda_e^- = \mu_{\rm eff} + 4 p_2^e B''(0) - \mu_c \quad , \quad
p_2^e = - \int_k \frac{1}{- t \Delta(k) + \mu_c}
\ee
which determines $\lambda_e$ as a function of $\mu_c$. It turns out (see below)
that this is always the {\it lower edge}, hence we denoted it $\lambda_e^-$.
We discuss below how to obtain the other edge(s) when they exist.

\subsubsection{Some examples: edges and the spectral density shape}

Let us study some examples, remembering that we denote
$\int_k = \frac{1}{L^d} \sum_k  \equiv \int \frac{d^d k}{(2 \pi)^d}$
for either discrete or continuum models.

\begin{enumerate}

\item
First recall that for a single-site (equivalently, zero dimensional $d=0$) system with $L^d=1$ the equation (\ref{diag2}) gives
\be
i p = \frac{1}{\lambda - \mu_{\rm eff} -  4 i p B''(0)} \quad  \Leftrightarrow \quad
i p = \frac{\lambda - \mu_{\rm eff} + i \sqrt{16 B''(0) -  (\lambda - \mu_{\rm eff})^2}}{ 8 B''(0)}
\ee
Hence the Hessian spectral density from (\ref{denim}) is given by the semicircular law
\be
\rho(\lambda) = \frac{1}{\pi} {\rm Im}(i p) = \frac{1}{ 8 \pi B''(0)}  \sqrt{16 B''(0) -  (\lambda - \mu_{\rm eff})^2}
~~  \theta_{\lambda}\left([\lambda_e^-,\lambda_e^+]\right)
\ee
\[
 \lambda_e^{\pm} = \mu_{\rm eff} \pm 4 \sqrt{B''(0)} \label{tt}
\]
where $\theta_{\lambda}([a,b])=1$ if $\lambda\in[a,b]$ and zero otherwise. This is precisely the result obtained in \cite{UsToy}. On the other hand we can determine the edge using \eqref{eliminate0}-\eqref{eliminate00}.
First let us examine the equation \eqref{Lark}. In that case it reads
\be
1 = \frac{4  B''(0)}{\mu_c^2}  \label{Lark2}
\ee
The positive root is $\mu_c= 2 \sqrt{B''(0)}$ and from \eqref{simple}
we find
\be
\lambda_e^- = \mu_{\rm eff} - 4  \frac{B''(0)}{\mu_c} - \mu_c = \mu_{\rm eff} - 4 \sqrt{B''(0)}
\ee
and recover the lower threshold \eqref{tt}. If we now use the {\it negative root} of \eqref{Lark2},
$\mu_c = - 2 \sqrt{B''(0)}$ we obtain instead $\lambda_e^+ =  \mu_{\rm eff} + 4 \sqrt{B''(0)}$,
i.e. the upper edge ! \\

It is easy to see that this is a general property. In other words the equation
\eqref{Lark} may have several roots. Let us call ${\cal D}$ the set on the real axis supporting the spectrum of
$- t \Delta$. It is easy to see that for the continuum
model, which has ${\cal D}=[0,+\infty[$, Eq. \eqref{Lark} may have only a single root.
In contrast, consider e.g. the infinite discrete lattice in $d=1$ with
$- t \Delta(k)=2 t (1-\cos k)$ so that its spectrum is in ${\cal D}=[0,4 t]$.
Clearly the r.h.s of \eqref{Lark} is infinite for $\mu_c \in - {\cal D} = [-4 t,0]$
and diverges at the edges of this interval. Hence one expects two roots,
one for $\mu_c=\mu^+_c>0$, and, by symmetry, one for $\mu_c= \mu_c^- = - 4 t - \mu^+_c$.
In the following we will always associate the positive root $\mu_c^+=\mu_c$ with the Larkin mass.
If the set $- {\cal D}$ consists of several intervals, or several points, where
the r.h.s. of \eqref{Lark} is infinite, there can be several additional solution
to \eqref{Lark} besides one associated with the Larkin mass. That one
we know must be the largest one since $- t \Delta$ is required to be positive
definite. Hence it corresponds to the lower edge of the Hessian. \\

\item
As soon as we have $L \geq 2$ the spectral density for the Hessian is not a semicircle
as we now discuss.
For a line $d=1$ with $L$ points and periodic boundary conditions
the eigenvalues of $- t \Delta$ are
$2 t (1- \cos \frac{2 \pi j}{L})$, $j=1,\dots,L-1$. The equation
\eqref{Lark} becomes
\be\label{simpleA}
\sum_{j=0}^{L-1} \frac{1}{(\mu_c + 2 t (1- \cos \frac{2 \pi j}{L}))^2} = \frac{L}{4 B''(0)}
\ee

It is easy to see that for very weak disorder, $\frac{B''(0)}{t^2 L} \ll 1$, there are $2L$ roots to this equation,
which we denote
$\mu_c=\mu_c^{j,\pm}$, $j=0,\dots,L-1$ with
\bea
\frac{1}{t} \mu_c^{j,\pm} = - 2 (1- \cos \frac{2 \pi j}{L}) \mp 2 \sqrt{\frac{B''(0)}{t^2 L}} (1+ O(\frac{B''(0)}{t^2 L}))
\eea
which can be found by considering successively all the quadratic divergences in each term in the sum in the left-hand side of  \eqref{simpleA} and approximating the sum accordingly. The formula \eqref{simple} for the corresponding edge becomes
\bea \label{simple2}
\lambda_e = \mu_{\rm eff} - \mu_c -  \frac{4 B''(0)}{L}
\sum_{j=0}^{L-1} \frac{1}{\mu_c + 2 t (1- \cos \frac{2 \pi j}{L})}
\eea
Substituting here the values of $\mu_c^{j,\pm}$ found above we
arrive, up to subdominant terms at weak disorder, to the corresponding edge values
\bea \label{bands}
\lambda_e^{j,\pm}  = \mu_{\rm eff} +
2 t (1- \cos \frac{2 \pi j}{L}) \pm 4 \sqrt{\frac{B''(0)}{L}} + \dots
\eea
For $L=1$ the above approximation is exact and one recovers the formula \eqref{tt}
valid for any disorder. For $L \geq 2$ there are $2 L$ edges and $L$ bands at weak
disorder. It is easy to see why. When disorder is zero the Hessian is simply the Hessian of
the elastic matrix and its spectrum is the set of delta peaks at
$2 t (1- \cos \frac{2 \pi j}{L}) + \mu$ (in that case $\mu_{\rm eff}=\mu$).
As disorder increases, each of these delta peaks broadens,
leading to a band, as described by \eqref{bands}.
One can expect that these bands will remain well separated
as long as their width $8 \sqrt{\frac{B''(0)}{L}}$ is much smaller
than their separations $\approx \frac{4 t}{L}$. This gives the
criterion
\be
\frac{4 L B''(0)}{t^2} \ll 1
\ee
to have separated bands. It is reasonable to expect that
in that situation each band will have a semi-circle form,
since each basically solves independently the $d=0$ equation.

To study the merging of such bands, let us consider the case
$L=2$ in more details (the eigenmodes are then $k=0,\pi$). The equation
becomes
\be
\mu_c = - 2 t + 2 t z \quad , \quad y = \frac{2 t^2}{B''(0)}  \quad , \quad \frac{1}{(1-z)^2} + \frac{1}{(1+z)^2} = y
\ee
Hence there are two cases. Either disorder is weak $\frac{2 t^2}{B''(0)} > 2$,
and there are 4 real roots
\be
z_{\pm,+} = \pm \sqrt{\frac{y+\sqrt{4 y+1}+1}{y}}  \quad , \quad z_{\pm,-} = \pm \sqrt{\frac{y-\sqrt{4 y+1}+1}{y}}
\ee
with $|z_{\pm,+}|>1$ and $|z_{\pm,-}|<1$ always. Or
disorder is strong and only the two roots $z_{\pm,+}$ exist. These roots correspond to edges
of the spectrum of the Hessian
\be
\lambda_e = \mu_{\rm eff} - \mu_c -  4 B''(0) \int_k \frac{1}{- t \Delta(k) + \mu_c}
= \mu_{\rm eff} + \frac{B''(0)}{t} [ y (1-z) - \frac{1}{z-1} - \frac{1}{z+1})
\ee
\be
 = \mu_{\rm eff} + \frac{B''(0)}{t} [  y \mp \frac{1}{2} z_{+,\epsilon} (2 y - 1 + \epsilon \sqrt{1+4 y})]
\ee
There are thus 4 edges (weak disorder) and 2 edges (strong disorder). The lowest edge is
located at
\bea
\fl && \lambda_e^{-} = \mu_{\rm eff} + \frac{B''(0)}{t} [  y - \frac{1}{2} z_{++} (2 y - 1 +  \sqrt{1+4 y})]
 \\
 \fl && = \mu_{\rm eff} + \frac{B''(0)}{4 t} \left(w^2-\sqrt{w-1} (w+3)^{3/2}-1\right), \quad
w = \sqrt{1 + 4 y} = \sqrt{ 1 + \frac{8 t^2}{B''(0)} } \nonumber
\eea

We can now study the spectral density for the case $L=2$.  It is given by $\rho(\lambda) = \frac{1}{\pi} {\rm Im}(i p)$
where $ip$ satisfies the cubic equation
\be
2 i p =  \frac{1}{\lambda -\mu_{\rm eff}  - 4 i p B''(0)} + \frac{1}{\lambda -\mu_{\rm eff}  - 4 t - 4 i p B''(0)}
\ee
The resulting density of states is plotted in Fig. \ref{fig4a} and \ref{fig4b}. The
evolution described above from disjoint supports (weak disorder) to a single
support (strong disorder), as well as the transition at $B''(0)/t^2=1$ is
clearly visible.

\begin{figure}[h!]
\centering
\includegraphics[width=0.49\textwidth]{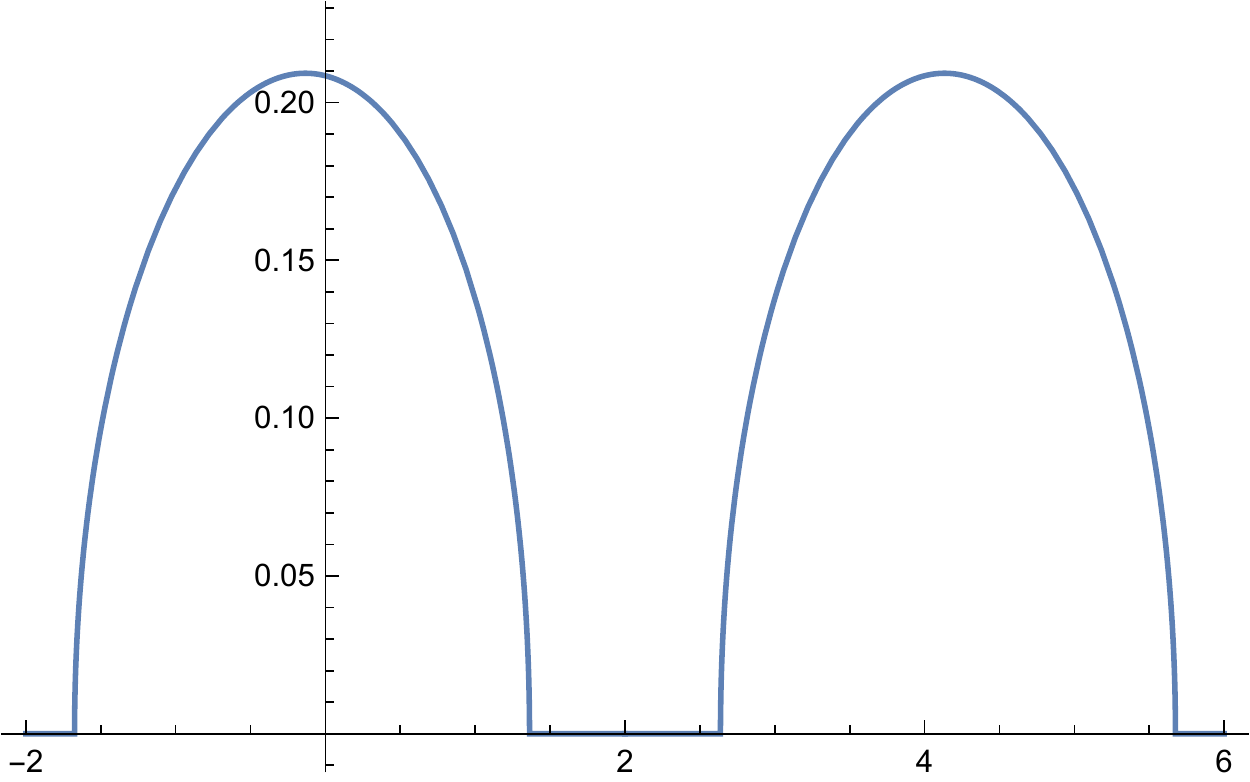}
\includegraphics[width=0.49\textwidth]{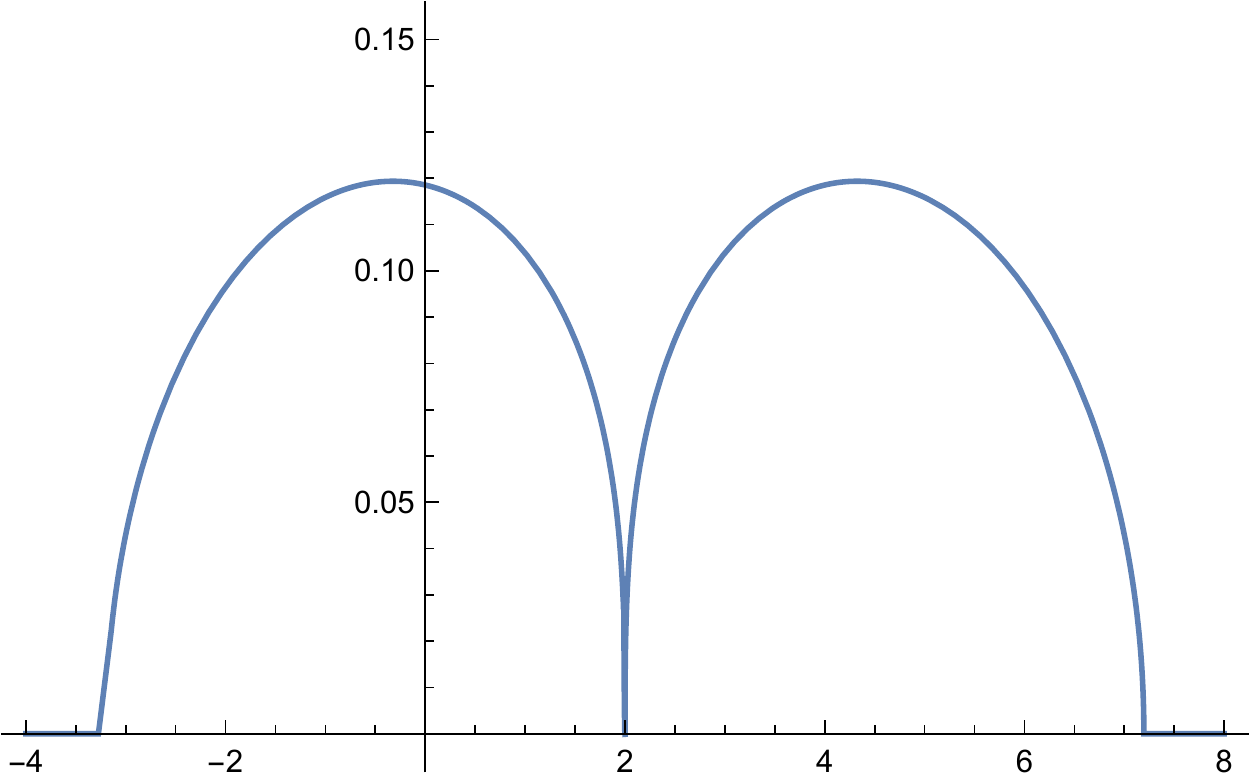}
\caption{ Density of states $\rho(\lambda)$ for two coupled points, $L=2$. We set $t=1$. Left:
weak disorder $B''(0)=0.3$, the support splits in two disjoint intervals where the density of states is close to a semi-circle
as for $L=1$. Right: critical disorder $B''(0)=1$ at which the two interval touch.}\label{fig4a}
\end{figure}

\begin{figure}[h!]
\centering
\includegraphics[width=0.49\textwidth]{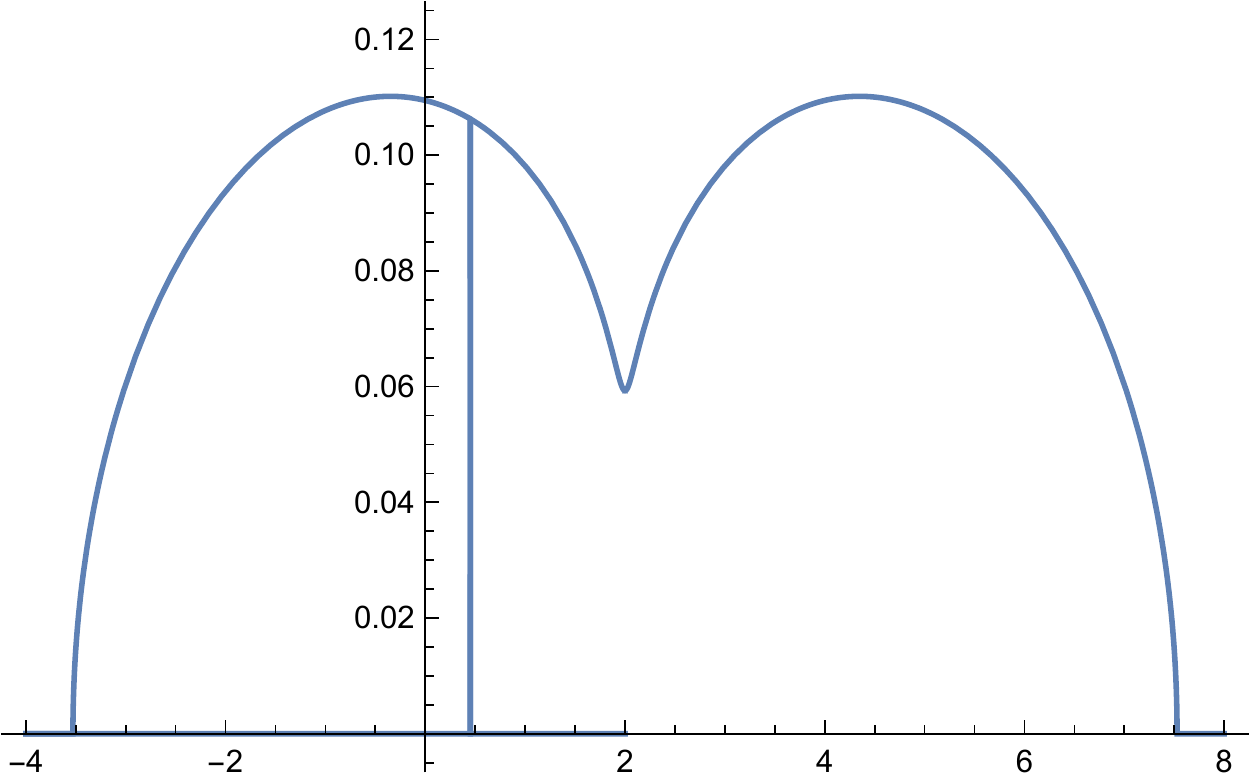}
\includegraphics[width=0.49\textwidth]{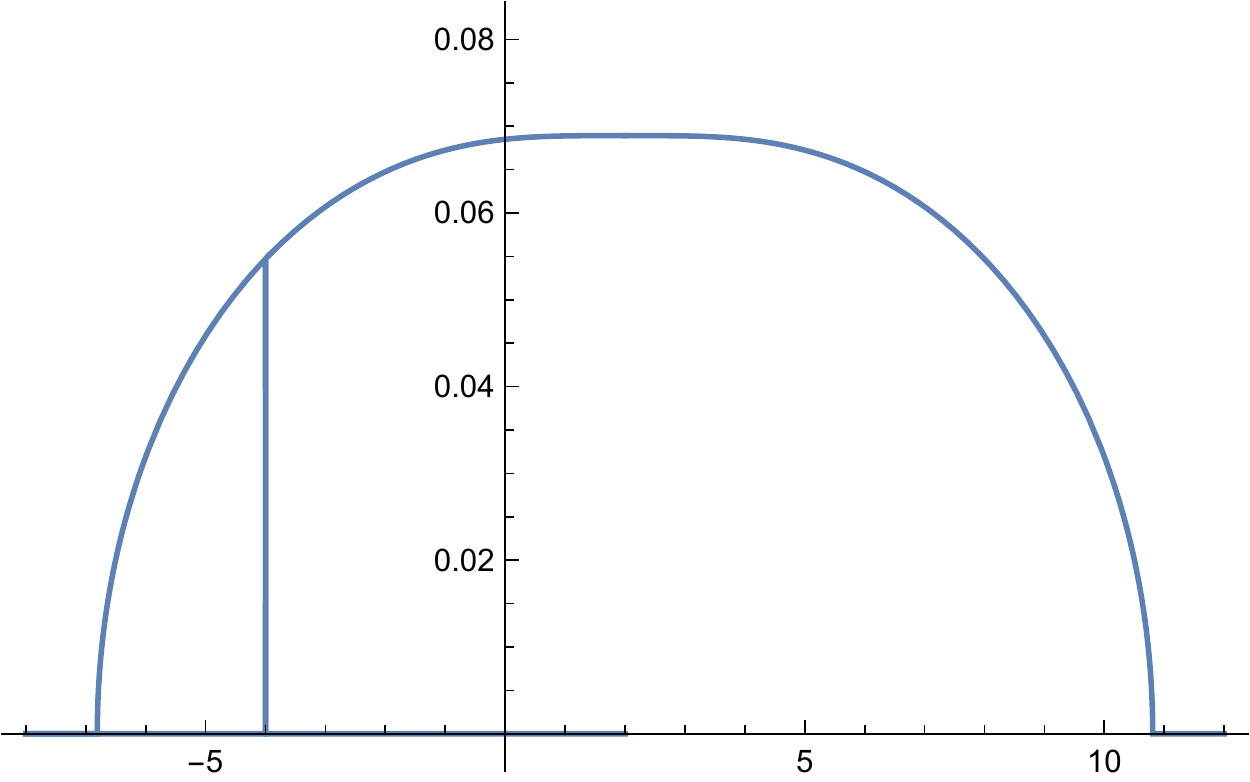}
\caption{Density of states $\rho(\lambda)$ for two coupled points, $L=2$. We set $t=1$. Left:
$B''(0)=1.2$. Right: larger disorder $B''(0)=4$. The vertical bar arises from a switch in the solution of the cubic equation.}\label{fig4b}
\end{figure}

\item
Our next example is the continuum $1D$ line of infinite length $L \to +\infty$.
Since the Laplacian spectrum for such a system is given by $- t \Delta(k)=t k^2$ with $k\in [0,+\infty[$
for such a case there is only one spectral edge in the system with disorder, the lower edge $\lambda_e^-$, determined from the Larkin mass
$\mu_c$, i.e. the unique positive solution of
\be
\frac{1}{4 B''(0)} = \int_{-\infty}^{\infty} \frac{dk}{2 \pi} \frac{1}{(t k^2 + \mu_c)^2}
= \frac{1}{4 t^{1/2} \mu_c^{3/2}}
\ee
leading to
\be\label{1dboundary}
\mu_c= (\frac{B''(0)}{\sqrt{t}})^{2/3} \quad , \quad
\lambda_e^{-} = \mu_{\rm eff} - \mu_c -  2 \frac{B''(0)}{\sqrt{t}}  \mu_c^{-1/2} =
\mu_{\rm eff}  - 3 (\frac{B''(0)}{\sqrt{t}})^{2/3}
\ee
The spectral density in the interval $\lambda>\lambda_e^{-}$ is then given by $\rho(\lambda) = \frac{1}{\pi} {\rm Im}(i p)$ where the complex $p$ is obtained by solving the following equation:
\bea\label{contcase1}
 i p = \int_{-\infty}^{\infty} \frac{dk}{2 \pi} \frac{1}{\lambda -\mu_{\rm eff} - t  k^2 - 4 i p B''(0)}\,.
\eea

Introducing new scaled  variables $y, \tilde{p}, \tilde{\lambda}$ via
\be\label{scaledvar}
y = \frac{t^2}{B''(0)} , \quad  p=\frac{\tilde{p}}{2y^{1/6}\sqrt{B''(0)}}, \quad \lambda-\mu_{eff}=\frac{\tilde{\lambda}\sqrt{B''(0)}}{y^{1/6}}
\ee
and rescaling the integration variable as $k\to k\,y^{-1/3}$ the equation (\ref{contcase1}) takes the form
\be\label{contcase2}
 i \tilde{p} = \frac{1}{\pi}\int_{-\infty}^{\infty} \frac{dk}{\tilde{\lambda}- 2 i \tilde{p}- k^2 }=\frac{-i}{ \sqrt{\tilde{\lambda}-2i\tilde{p}}} \, \mbox{sgn} \left(\mbox{Im}\sqrt{\tilde{\lambda}-2i\tilde{p}}\right), \quad \! \mbox{Im}\sqrt{\tilde{\lambda}-2i\tilde{p}}\ne 0
\ee

Taking the square and
further introducing the variable $w$ and parameter $\delta$ by
\be\label{param}
\tilde{p}=-\frac{i}{w}, \quad \tilde{\lambda}=3\delta^{1/3}
\ee

the equation for $w$ attains especially simple form:
\be\label{scaledcubic}
w^3+3\delta^{1/3}w-2=0
\ee

The general theory of cubic equations then dictates that for $\delta<-1$ the equation $\eqref{scaledcubic}$ has only real solutions,
hence $p\sim -i w^{-1}$ will be purely imaginary implying zero density of eigenvalues. This parameter range fully agrees with the position of lower spectral threshold $\lambda_{e}^{-}$ found in (\ref{1dboundary}), which in the new variables reads $\tilde \lambda_{e}^{-}=-3 y^{-1/6}$. As long as $\delta>-1$ there is one real root
given by the Cardano formula in the form
\be\label{card1dcont}
w_r=\left[1+\sqrt{1+\delta}\right]^{1/3}+\left[1-\sqrt{1+\delta}\right]^{1/3}
\ee
which is positive and decreases from $2$ to $0$ as $\delta$ increases from $-1$ to $+\infty$. There are
also two complex conjugated solutions $w$ and $\overline{w}$. To find their imaginary part we use the Vieta's formulas:
\[
w_r+w+\overline{w}=0, \quad w_r \,w \overline{w}=2
\]
which gives $Re(w)=-\frac{w_r}{2}$ and  $Im(w)=\pm \sqrt{\frac{2}{w_r}-\frac{w_r^2}{4}}$. It is then easy to see that the spectral density can be found explicitly and is given by:
\[
\rho(\lambda)=\frac{1}{2\pi \sqrt{B''(0)}y^{1/6}}Im\left(w^{-1}\right)=-\frac{1}{2\pi \sqrt{B''(0)} y^{1/6}}\frac{Im(w)}{Re^2(w)+Im^2(w)}
\]
where we have to choose the sign of $Im(w)$ which ensures positivity of the mean density.
Recalling that in terms of the original variables
\[
\delta=\frac{t}{[B''(0)]^2}\frac{(\lambda-\mu_{eff})^3}{27}=\left(-1+\frac{\lambda-\lambda_e^{-}}{\mu_{eff}-\lambda_e^{-}}\right)^3, \quad y=\frac{t^2}{B''(0)}
\]
this finally implies
\be\label{den1dcontexplicit}
\rho(\lambda) = \frac{1}{2\pi (t\,B''(0))^{1/3}} ~r_c\left(\Lambda=t^{1/3} \frac{\lambda-\mu_{\rm eff}}{3 B''(0)^{2/3}}\right), \quad r_c(\Lambda)=\frac{w_r^2}{4}\sqrt{\left(\frac{2}{w_r}\right)^3-1}
\ee

We have plotted in the Fig. \ref{fig3} the parameter free scaling function $r_c(\Lambda)=\sqrt{U-U^4}, \, U=\frac{w_r}{2}$
with $w_r$ given by  \eqref{card1dcont} and $\delta=\Lambda^3$.
It has the
following asymptotics for large $\Lambda$ and for $\Lambda$ near the edge $\Lambda_e=-1$:
\bea
r_c(\Lambda\gg 1) &=& \frac{1}{\sqrt{3 \Lambda}}-\frac{5}{54
   \sqrt{3} \Lambda^{7/2}}+O( \Lambda^{-13/2} ) \label{largelambda}\\
r_c(\Lambda+1\ll 1)&=& \sqrt{\Lambda +1}-\frac{5}{18} (\Lambda
   +1)^{3/2}+O\left((\Lambda +1)^{5/2}\right)
\eea

  In particular, equation (\ref{den1dcontexplicit}) then implies
that
\be \label{rr}
\rho(\lambda) \simeq \frac{1}{2\pi (t\,B''(0))^{1/3}} \sqrt{\frac{\lambda-\lambda_e^{-}}{\mu_{\rm eff}-\lambda_e^{-}}}
\ee
showing the expected square-root  singularity close to the spectral edge $\lambda=\lambda_e^{-}$.
Moreover, it is easy to see that $\frac{dr_c}{dU}=0$ for $U=2^{-2/3}$, hence $w_r=2^{1/3}$ corresponding according to
\eqref{card1dcont} to $\delta=0$. We conclude that $\Lambda=0$ is exactly the point of the maximum for the scaled density profile $r_c(\Lambda)$, which is readily seen from Fig. \ref{fig3}.

\item
Our last example is an infinite discrete lattice $d=1$ with the number of sites $L \to \infty$.
The Laplacian spectrum for such a system is given by $- t \Delta(k)=2t (1-\cos{k})$ with $k\in[0,2\pi]$.
To determine the spectral edges we use the integrals
\begin{equation}\label{intreal1d}
 \int_0^{2 \pi} \frac{dk}{2 \pi} \frac{1}{\cos k+x} = \frac{{\rm sgn}(x)}{\sqrt{x^2-1}}, \quad \int_0^{2 \pi} \frac{dk}{2 \pi} \frac{1}{(\cos k+x)^2} = \frac{|x|}{(x^2-1)^{3/2}}
\end{equation}
 for real $|x|>1$.
This reduces finding the roots of \eqref{Lark} to solving the equation
\be
\frac{1}{4 B''(0)} = \int_0^{2 \pi} \frac{dk}{2 \pi} \frac{1}{(2 t (1-\cos k) + \mu_c)^2}
= \frac{|\mu_c + 2 t|}{(\mu_c (\mu_c + 4 t))^{3/2}}
\ee
for $\mu_c>0$ or $\mu_c<-4 t$. Denoting $r=1+\frac{\mu_c}{2t}$ and $ y = \frac{t^2}{B''(0)}$ we rewrite the above equation as
$\frac{|r|}{(r^2-1)^{3/2}} = y$ which implies a simple cubic equation $w^3-y^{-2/3}w-1=0$ for $w=(r^2-1)y^{2/3}$. Note that we have $|r|>1$ for all allowed choices of $\mu_c$, hence need to look for a real positive solution $w>0$ of this equation. According to general properties of cubic equations, for $y>\frac{2}{3\sqrt{3}}$ our equation has only a single real  root $w=w_c$ given by the Cardano formula:
\[
w_c=\Delta^{1/3}+\frac{1}{3y^{2/3}}\Delta^{-1/3}, \quad \Delta=\frac{1}{2}+\sqrt{\frac{1}{4}-\frac{1}{27y^2}}
\]
which is obviously positive as needed for our goals.
For the parameter $\mu_c$ we then have two solutions. The positive one corresponds to the Larkin length:
\be\label{Larkin1d}
\mu_c=\mu_c^{+}=2t\left(\sqrt{1+w_cy^{-2/3}}-1\right)>0
\ee
and the second solution $\mu_c^{-}=-2t\left(\sqrt{1+w_cy^{-2/3}}+1\right)\equiv - 4 t - \mu_c$ as expected by symmetry.

In the case $0\le y\le \frac{2}{3\sqrt{3}}$ the cubic equations has all three real roots. Introducing the angle $\theta\in [0,\frac{\pi}{2}]$ such that $\cos{\theta}=\frac{3\sqrt{3}}{2}y$ the roots can be conveniently written in the so-called trigonometric form:
\[
w_c^{(1)}=\frac{2}{\sqrt{3}y^{1/3}}\cos{\left(\frac{\theta}{3}\right)}>0, \quad w_c^{(2)}=\frac{2}{\sqrt{3}y^{1/3}}\cos{\left(\frac{\theta+2\pi}{3}\right)}<0
\]
\[
w_c^{(2)}=\frac{2}{\sqrt{3}y^{1/3}}\cos{\left(\frac{\theta+4\pi}{3}\right)}<0,
\]
so only $w_c^{(1)}$ can be used for the above procedure and yields $\mu_c^{\pm}$.

Finally, this gives us the two spectral edges as
\be
\lambda_e^{\pm} = \mu_{\rm eff} - \mu^{\mp}_c -  4 B''(0) \int_k \frac{1}{2 t (1- \cos k) + \mu^{\mp}_c}
= \mu_{\rm eff} - \mu_c^{\mp}- 4 B''(0)\frac{{\rm sgn}\left(\mu_c^{\mp}+2t\right)}{\sqrt{\mu_c^{\mp} (4 t+\mu^{\mp}_c)}}
\ee

Let us give a simple example: $t=2^{1/2}, B''(0)=3^{3/2}$ which gives $y=\frac{2}{3^{3/2}}$,
hence $\Delta=1/2$ and $w_c=2^{2/3}$
which eventually gives for the Larkin length $\mu_c^{+}=2\sqrt{2}$ and $\mu_c^{-}=-6\sqrt{2}$.

To calculate the spectral density profile one needs the following generalization of (\ref{intreal1d}):
\begin{equation}\label{intcomp1d}
  \frac{1}{2 \pi}\int_0^{2 \pi} \frac{dk}{2t\cos k+a} = \frac{{\rm sgn}(|a+\sqrt{a^2-4t^2}|-2t)}{\sqrt{a^2-4t^2}}
\end{equation}
valid for any real $t>0$ and complex $a$
 such that $|a+\sqrt{a^2-4t^2}|-2t\ne 0$, excluding $a$ real in the interval $[-2t,2t]$.

The spectral density  in the interval $\lambda_e^{-}<\lambda<\lambda_e^{+}$ is then given by $\rho(\lambda) = \frac{1}{\pi} {\rm Im}(i p)$ where the complex $p$ is obtained by solving the following equation, cf. (\ref{diag2}):
\bea \label{sc1}
  i p &=& \int_0^{2 \pi} \frac{dk}{2 \pi} \frac{1}{\lambda -\mu_{\rm eff} + 2 t  (\cos k -1) - 4 i p B''(0)}
\\
&=& \frac{{\rm sgn}(|a(p)+\sqrt{a^2(p)-4t^2}|-2t)}{\sqrt{a^2(p)-4t^2}}
\eea
where we denoted $  a(p)=\lambda-2 t - \mu_{\rm eff}  - 4 i p B''(0) $. We define the scaled variables
\be
y = \frac{t^2}{B''(0)} \quad , \quad
\lambda- \mu_{\rm eff} = 2 t \Lambda \quad , \quad
p = \frac{\sqrt{y}}{2 \sqrt{B''(0)}}  P
\ee
which brings Eq. \eqref{sc1} in the dimensionless form
\be \label{eq11}
i P  y = \frac{s}{\sqrt{ (\Lambda - 1 - i P)^2 -1}}
\ee
where $s=\pm 1$ with $s={\rm sgn}( |\Lambda -1 - i P + \sqrt{ (\Lambda - 1 - i P)^2 -1}|-1)$.
The roots of \eqref{eq11} must satisfy the following equation
for  $\tilde{P}=iP$
\be \label{quartic}
\tilde{P}^4+2(1-\Lambda)\tilde{P}^3-(2-\Lambda)\Lambda \tilde{P}^2-\frac{1}{y^2}=0
\ee

The spectral density is then given in terms of the function  $ r(\Lambda,y)={\rm Im}(\tilde{P})$ as
\be \label{dens1}
\rho(\lambda) =  \frac{t}{2 \pi B''(0)} ~ r\left(\Lambda=\frac{\lambda- \mu_{\rm eff} }{2 t},y\right) \, ,
\ee

The parameter $y$ reflects the strength of the disorder relative to the elasticity, so that
the larger is $y$ the weaker is the disorder.
For strong disorder (or vanishing elasticity), $y\to 0$, the system decouples in
non-interacting zero-dimensional units and the spectral density is given by the semicircular law,
as can be seen e.g. setting $t=0$ in \eqref{sc1}.
For a moderate disorder $y\sim 1$ the shape is not a semicircle any longer, but is qualitatively similar, as
can be seen in Fig. \ref{fig1} where the scaling function $r(\Lambda,y)$ is plotted for $y=1$.
However with decreasing disorder/increasing elasticity the shape
of the spectral density changes qualitatively and develops a characteristic form with two maxima and a minimum
in between, see plot for $y=10$ in Fig.\ref{fig2}.

Some hints towards the origin of such shape can be obtained by considering
the limit of vanishing disorder $B''(0) \to 0$, i.e. $y \to +\infty$. In this limit one expects that the
spectral density should in a certain sense converge to the one of the purely elastic $1d$ system:
\[
\rho(\lambda) = \int_0^{2 \pi} \frac{dk}{2 \pi} \delta( \lambda -\mu_{\rm eff} + 2 t  (\cos k -1) )
\]
\be\label{pureelasticlattice}
= \frac{1}{2 t} \int_0^{2 \pi} \frac{dk}{2 \pi} \delta( \Lambda - (1-\cos k) )
 = \frac{1}{2 \pi t} \frac{1}{\sqrt{\Lambda (2-\Lambda)} },
\ee
which implies that for  $y\gg 1$, $r(\Lambda,y) \simeq 1/(y \sqrt{\Lambda (2-\Lambda)})$.
 The correspondence with the disorder-free result is visible on the Fig. \ref{fig2}. in the central part around  the minimum.

 To understand the two-maxima shape we investigate analytically the case of large but finite $y\gg 1$ more accurately.
 Rescaling $\tilde{P}=q/y$ the equation (\ref{quartic}) takes the form:
 \be \label{quarticscaled}
\frac{1}{y^2}q^4+\frac{2}{y}(1-\Lambda)q^3-(2-\Lambda)\Lambda q^2-1=0
\ee
 Now it is obvious that letting $y\to \infty$ for a fixed $0<\Lambda<2$  the above is reduced to the quadratic equation
  with purely imaginary roots $q=\pm \frac{i}{\sqrt{(2-\Lambda)\Lambda)}}$. This solution yields precisely the
  density for the pure elastic case (\ref{pureelasticlattice}). However, it is also evident that in the vicinity of the points $\Lambda=0$ or $\Lambda=2$ such naive limit breaks down and requires a separate treatment. We illustrate it
  by providing analysis in the vicinity of $\Lambda=0$, one for the region around $\Lambda=2$ being fully analogous. A simple scaling argument demonstrates that
  the relevant vicinity of $\Lambda=0$ is of the width $|\Lambda|\sim y^{-2/3}$ so that it makes sense to introduce a new parameter
  $\delta=\left(\frac{2}{3}y^{2/3}\Lambda\right)^3$ and also introduce the scaled variable $w$ via $q=\frac{y^{1/3}}{w}$.
  Substituting this into  (\ref{quarticscaled}) and taking the limit $y\to \infty$ one finds the equation for $w$ exactly given by the equation   (\ref{scaledcubic}) studied by us in much detail above in our analysis of the $1d$ disordered continuum problem.
  We therefore conclude that for $y \to \infty$ and around $\Lambda=0$ the scaled spectral density profile $r(\Lambda,y)$ for the $1D$ discrete model is simply given in terms of the scaled density profile of the continuum model $r_c(\Lambda)$ obtained in \eqref{den1dcontexplicit}
  as $r(\Lambda,y)=y^{-2/3}r_c\left(\frac{2}{3}y^{2/3}\Lambda\right)$. In particular, recalling that $r_c(\Lambda=-1)=0$ in the continuum case, we find that
  the position of the left spectral threshold in the discrete case for $y>>1$ is given by $\Lambda_e^{-}=-\frac{3}{2}y^{-2/3}$.
  Close to this threshold the density increases as the square root $\sqrt{\Lambda-\Lambda_e^{-}}$, eventually reaching its maximal value
  $r(0,y)=y^{-2/3}2^{-4/3}\sqrt{3}$ exactly at $\Lambda=0$  and then decaying for larger $\Lambda>>y^{-2/3}$ in agreement with the asymptotics (\ref{largelambda}) as
  \[
  r(\delta\gg 1)\sim y^{-2/3} \frac{1}{\sqrt{3\left(\frac{2}{3}y^{2/3}\Lambda\right)}}= \frac{1}{y}\frac{1}{\sqrt{2\Lambda}}
  \]
  which precisely matches the $\Lambda<<1$ behaviour from the 'central part'  $r(\Lambda,y) \simeq 1/(y \sqrt{\Lambda (2-\Lambda)})$.
   This demonstrates that for weak disorder, $y\gg 1$, the shape of the spectral density for the $1D$ infinite elastic lattice is given by (i) a central part which converges to the pure, ''disorder-free'', density of states
  (ii) two edge regions, $|\Lambda|\sim y^{-2/3}$ and $|2-\Lambda|\sim y^{-2/3}$,
  where the divergent density of states of the pure system is converted into a finite profile, identical
  upon rescaling
  to the one of the $1D$ disordered continuous elastic line.

\begin{figure} h!
\centering
\includegraphics[scale=.8]{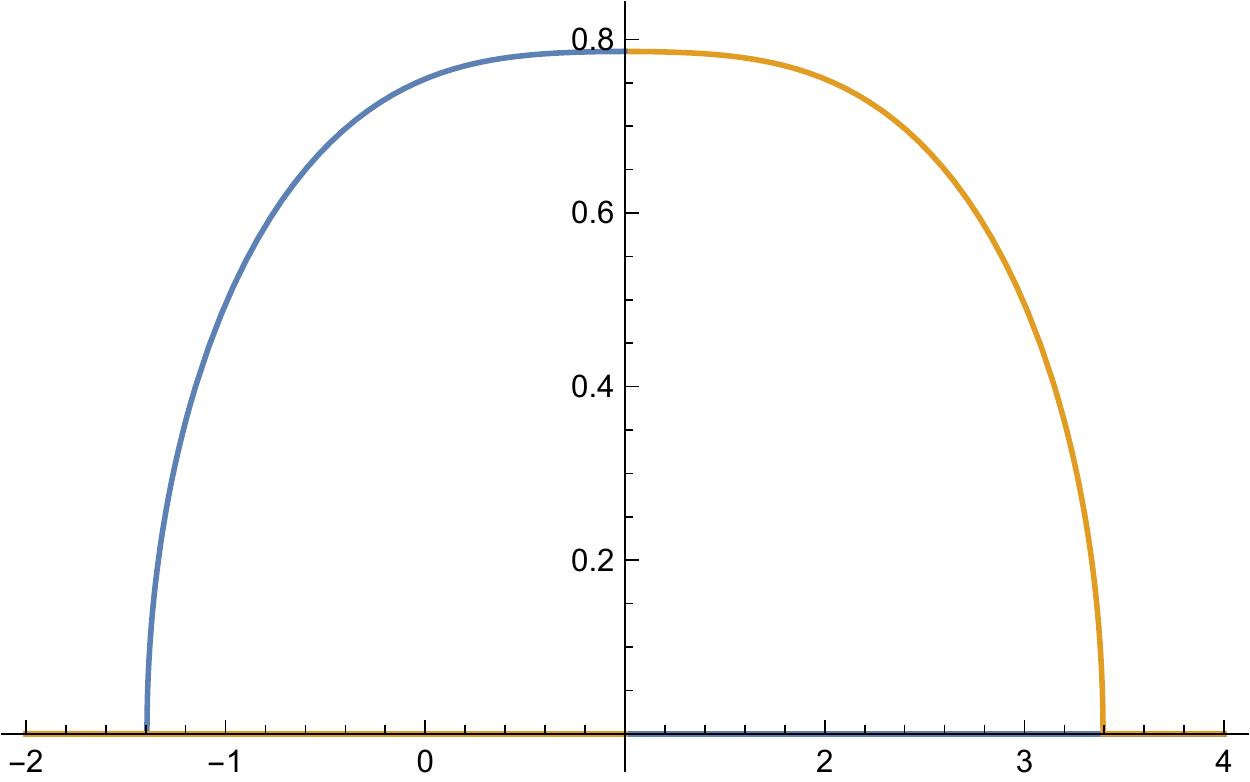}
\caption{Scaling function for the Hessian spectral density, $r(\Lambda,y)$ versus $\Lambda=\frac{\lambda- \mu_{\rm eff} }{2 t}$ given
by Eq. \eqref{dens1},
for  intermediate disorder $y=\frac{t^2}{B''(0)} = 1$. The two colors correspond
to two roots of the quartic equation \eqref{quartic}.}
\label{fig1}
\end{figure}

\end{enumerate}

\subsection{Phases from replica, determination of $\mu_{\rm eff}$ and of the gap}

\label{sec:phases}
In the previous section we have obtained the spectral density  and its support, in particular the lower edge,
for various cases. The formulas however contained a single as yet unknown parameter $\mu_{\rm eff}$
which corresponds to a global shift of the support of the Hessian spectral density.

In this section our aim is to determine $\mu_{\rm eff}$, the missing information about the
global position of the Hessian spectrum. As $\mu$ is varied the system
can be in different phases (RS, 1RSB, FRSB) and the formula leading
to $\mu_{\rm eff}$ must be detemined accordingly. In each
case we first recall briefly the known replica saddle point
solutions for the random manifold problem \cite{MP2,LDMW}, i.e. the solutions
to the equations \eqref{repsp1}-\eqref{repsp2}.

\subsubsection{Replica-symmetric phase}

Let us start with the replica symmetric (RS) phase, which occurs for $\mu>\mu_c$.
Let us look for a replica symmetric solution of the saddle point equations \eqref{repsp1}-\eqref{repsp2}.
\bea
&& \sigma_{ab}= \sigma_c \delta_{ab} + \sigma \quad ,  \quad  G_{aa}(k) = \tilde G(k) \quad , \quad G_{a \neq b}(k) = G(k)
\eea
Note that the condition $\sum_b\sigma_{ab}=0$ in this parametrization reads $\sigma_c+n\sigma=0$ which in the replica limit $n\to 0$ implies that we can choose $\sigma_c=0$. We also have $\left[G^{-1}\right]_{a\ne b}=-\sigma,  \left[G^{-1}\right]_{aa}=\mu-t\Delta(k)-\sigma(1-n)$ and
the inversion of the RS matrix gives:
\be
 G_{aa}(k) = \frac{\mu-t\Delta(k)-\sigma-\sigma(n-2)}{\left(\mu - t \Delta(k)\right)\left(\mu - t \Delta(k)-\sigma n\right)}, \quad G_{a\ne b}(k) = -\frac{- \sigma}{\left(\mu - t \Delta(k)\right)\left(\mu - t \Delta(k)-\sigma n\right)},
\ee
or equivalently in the replica limit $n\to 0$
\be
 G_{ab}(k) =  \frac{1}{\mu - t \Delta(k)} \delta_{ab} + \frac{\sigma}{(\mu - t \Delta(k))^2}
\quad , \quad \chi_{a \neq b}= 2 T  \int_k \frac{1}{\mu - t \Delta(k)}
\ee
with $\chi_{aa}=0$ by definition. The saddle point equation \eqref{repsp2} leads then to the explicit formula for
$\sigma$, which determines completely the solution
\be
 \sigma = - 2 \beta B'\left(2 T \int_k \frac{1}{\mu - t \Delta(k)}\right)
\ee
and $Q_{ab} =  T \int_k G_{ab}(k)$. As is well known \cite{MP2}
the RS solution is valid for $\mu > \mu_c(T)$ with
(see Eq. (17) in \cite{LDMW})
\be\label{stabRS}
1 = 4  \int_k \frac{1}{(\mu_c(T) - t \Delta(k))^2} B''\left(2 T \int_k \frac{1}{\mu_c - t \Delta(k)}\right)
\ee
which gives, in the $T=0$ limit, $\mu_c(T=0)=\mu_c>0$, i.e. the Larkin mass determined by
\be\label{Larkinmass}
1 = 4 B''(0)  \int_k \frac{1}{(\mu_c - t \Delta(k))^2}
\ee
as anticipated in the previous Section.\\

We can now determine $\mu_{\rm eff}$ and the edges of the Hessian in the RS phase.
From \eqref{mueffdef} we obtain (for $n=0$)
\[
\mu_{\rm eff}=\mu- \frac{2}{T} \left[B'(\chi_{11})+(n-1)B'(\chi_{a \neq b})\right]\, \to\,
\mu- \frac{2}{T} \left[B'(0)-B'\left(2 T  \int_k \frac{1}{\mu - t \Delta(k)}\right)\right]
\]
\be\label{muefRS}
\simeq_{T \to 0} \mu + 4 B''(0)  \int_k \frac{1}{\mu - t \Delta(k)}
\ee

Substituting this value of $\mu_{\rm eff}$ in \eqref{simple} we thus obtain the final formula for
the {\it lower} spectral edge $\lambda^{(-)}_e$  of the Hessian (which we associate with the spectral gap)
as a function of $\mu$ in the RS phase
\bea \label{edgelow}
\lambda^{(-)}_e =  \mu - \mu_c
+ 4 B''(0)  \int_k \left[ \frac{1}{\mu - t \Delta(k)}  - \int_k \frac{1}{\mu_c - t \Delta(k)} \right]
\eea
This formula immediately shows that the gap vanishes {\it quadratically} at $\mu_c$, i.e. upon expanding
for $\mu>\mu_c$
\bea
\lambda^{(-)}_e =
4 B''(0)  \int_k  \frac{1}{(\mu_c - t \Delta(k))^3} \, (\mu-\mu_c)^2 + O((\mu-\mu_c)^3)
\eea
where the linear term cancels as a consequence of \eqref{Larkinmass}.
In $d=0$ we recover the similar formula obtained in \cite{UsToy}.

For the continuum model $- t \Delta(k)=t k^2$ there is only one edge, the lower edge,
which we just determined.
However, as extensively discussed in the previous Section, for other models (e.g.
discrete models) there may be several edges (and bands). As discussed there
extensively all the edges $\lambda_e^\alpha$ are obtained by considering
all the real roots $\mu_c^{\alpha}$ of \eqref{Larkinmass} and inserting them in the formula \eqref{edgelow}.
We refer to that Section for details. Let us simply note the case of the discrete $d=1$
model, where by symmetry the two roots of \eqref{Larkinmass} are $\mu_c$ and $-4 t - \mu_c$.
That gives the upper edge for that model, and one can then write a single formula for
both edges
\be
\lambda^{(\pm),{\rm 1d,discrete}}_e =
\mu + 2 t  + 4 B''(0)  \int_k  \frac{1}{\mu - t \Delta(k)} \pm \left(
2 t + \mu_c + 4 B''(0)  \int_k  \frac{1}{\mu_c - t \Delta(k)} \right)
\ee
which gives simple formula for the midpoint and the band width.

\subsubsection{Full RSB phase}

Let us now discuss the full RSB solution. We choose not to give the finite-$n$ hierarchical structure here as it is relatively
cumbersome, but rather simply follow the $n\to 0$ analysis
\footnote{in particular the Section VIII B of
\cite{PLDKWLargeNDetails} (note the small misprint in the definition
of $\sigma$ at the very beginning of the section there, paragraph above (8.4): it should be $T G^{-1}_{ab}(k)- (k^2+m^2) \delta_{ab} = - \sigma_{ab}$). Note that $f=\hat f= - B$ in \cite{MP2} and we follow \cite{MP2} for the definition of $G(k)$ i.e. we do not absorb the $T$ inside it as
done in \cite{LDMW,PLDKWLargeNDetails}.}
in \cite{MP2,TGPLDBragg1a,PLDKWLargeNDetails,LDMW}.
The off-diagonal part $\sigma_{a \neq b}$ is represented by a Parisi function
$\sigma(v)$, $v \in [0,1]$, which usually is $\sigma(v)=\sigma_c$ for $v \in [1,v_c]$,
varies continously with $v$ for $v \in [v_\mu,v]$ and is constant for $0<v<v_\mu$.
It is determined by
\be \label{e1}
 \sigma(v) = - \frac{2}{T} B'\left(2 T \int_k (\tilde G(k) - G(k,v))\right)
\ee
where from RSB replica matrix inversion one has
\be \label{e2}
\tilde G(k) - G(k,v) =  \int_k \frac{1}{\mu - t \Delta(k) + \Sigma_c}
+ \int_v^{v_c} dw \frac{ \sigma'(w)}{(\mu - t \Delta(k) + [\sigma](w))^2}
\ee
where $[\sigma](v)$ is defined by
\be\label{sigmabr}
[\sigma](v) = v \sigma(v) - \int_0^v dw \sigma(w)  \quad , \quad [\sigma](v_c)=\Sigma_c
\ee

Taking in  \eqref{e1} the derivatives w.r.t. $v$ and exploiting \eqref{e2}, one finds that
for any interval of $v$ either (i) $\sigma(v)$ is constant
or (ii) it satisfies the {\it marginality condition}
\be \label{marg}
1 = 4  B''\left(2 T \int_k (\tilde G(k) - G(k,v))\right)\,\int_k \frac{1}{(\mu - t \Delta(k) + [\sigma](v))^2}
\ee
In particular at the breakpoint $v=v_c$ one has
\be
1 = 4
B''\left(2 T \int_k \frac{1}{\mu - t \Delta(k) + \Sigma_c} \right)\,\int_k \frac{1}{(\mu - t \Delta(k) + \Sigma_c)^2}
\ee
which by comparison with RS stability condition (\ref{stabRS}) implies that
\be
\Sigma_c(T)= \mu_c(T)- \mu
\ee
and at $T=0$, as a function of $\mu_c$ the Larkin mass determined by \eqref{Larkinmass}, one has
\be
\Sigma_c = \mu_c - \mu\,.
\ee

For use here and in the next Section let us define the notations for $l \geq 1$
\bea \label{In}
I_{l}(x) =  \int_k \frac{1}{(- t \Delta(k) + x)^l}  \quad , \quad I'_{l}(x) = - l I_{l+1}(x)
\eea
and somewhat abusively
\bea \label{I0}
I_0(x) = \int_k \log(- t \Delta(k) + x)   \quad , \quad I'_0(x)=I_1(x)
\eea

One can calculate the solution for $[\sigma](v)$  for arbitrary covariance $B$
(see e.g. formula (8.16) in \cite{PLDKWLargeNDetails}) as we now show.
We assume that $B''$ is a monotonous decreasing function ($B''' <0$).
Inverting the marginality condition \eqref{marg} and inserting in \eqref{e1}
leads to
\be
 \sigma(v) = - \frac{2}{T} B'\left((B'')^{-1}\left( \frac{1}{4 I_2\left(\mu+[\sigma](v)\right)}\right)\right)
\ee
Taking a derivative of the above w.r.t. $v$ with the help of the identities
\[
\frac{d[\sigma](v)}{dv}=v\sigma'(v), \quad \frac{d}{dv}f^{-1}\left(\phi(v)\right)=\frac{\phi'(v)}{f'\left(f^{-1}\left(\phi(v)\right)\right)}
\]
where $f^{-1}$ is the functional inverse of $f$,  we obtain after rearranging and assuming $ \sigma'(v)\ne 0$ the formula
\bea \label{marg1}
v= - 4 T \frac{I^3_2\left(\mu+ [\sigma](v)\right)}{I_3\left(\mu+ [\sigma](v)\right)} B'''\left((B'')^{-1}\left( \frac{1}{4 I_2\left(\mu+ [\sigma](v)\right)}\right)\right)
\eea
which determines by inversion $[\sigma](v)$ as a function of $v$, which must be an increasing
function. It is now convenient to introduce
\[
A(v)=(B'')^{-1}\left( \frac{1}{4 I_2\left(\mu+ [\sigma](v)\right)}\right), \quad \Rightarrow \quad  I_2\left(\mu+ [\sigma](v)\right)=\frac{1}{4B''(A(v))}
\]
and define the function $F(b)$ via the relation
\be\label{Funf}
\frac{1}{I_3(x)} = F\left( \frac{1}{4 I_2(x)} \right)
\ee
As the result, the relation (\ref{marg1}) can be rewritten as
\[
v=-\frac{T}{16}\frac{B'''(A(v))F\left(B''(A(v))\right)}{\left[B''(A(v))\right]^3}
\]

 Taking yet another derivative w.r.t. $v$ and noticing that $\frac{dA}{dv}<0$, this leads to the following condition for FRSB solution to exist
\be \label{condfrsb}
\frac{d}{dq}  \frac{B'''(q) F\left(B''(q)\right)}{\left(B''(q)\right)^3}  >0
\ee
where we used that $dq/dv <0$.
More precisely the condition for FRSB to hold in an interval $[v_\mu,v_c]$
is that \eqref{condfrsb} holds for $q \in [q(v_c),q(v_\mu)]$.

One may now notice that for a $d-$dimensional continuum model with Laplacian spectrum $-t\Delta(k)=t\,{\bf k}^2$ and
$dk\sim |{\bf k}|^{d-1} d|{\bf k}|$ the behaviour of the integrals $I_{l>d/2}(x)$ in \eqref{In} for $x\to 0$ is dominated by the infrared ($|{\bf k}|\to 0$) limit
and is given by  $I_{l>d/2}(x)\sim x^{-\left(l-\frac{d}{2}\right)}$. Taking $d<4$ we then see that (\ref{Funf}) implies
 in the limit of small $\mu$ and small $b$ the behaviour
\be \label{smallarg}
F(b) \sim b^\frac{6-d}{4-d}
\ee
 The same behavior also holds for discretized models
on an infinite $d$ dimensional lattice. Replacing $ F\left(B''(q)\right)$ in (\ref{condfrsb}) with the small-argument asymptotics
(\ref{smallarg}) leads then to the full-RSB condition, which we gave in the Introduction, see
(\ref{criterionintro}).

Note that the free energy fluctuation exponent
$\theta=\theta_F=d-2 + \frac{4-d}{1+\gamma} >0$ and the FRSB self-energy behaves as
$[\sigma](v) \sim v^{2/\theta}$
at small $v$ (for $\mu=0$).

Let us now calculate $\mu_{\rm eff}$ for the FRSB solution. For this we first set $v=v_c$ in (\ref{e2}) and (\ref{sigmabr}) and get the relations
\be \label{e2vc}
 \tilde G(k) - G(k,v_c) =  \int_k \frac{1}{\mu - t \Delta(k) + \Sigma_c}, \quad
\Sigma_c = v_c \sigma(v_c) - \int_0^{v_c} dw \sigma(w)
\ee

Now inserting the
FRSB form into the definition \eqref{mueffdef} we get
\bea
 \mu_{\rm eff} &=& \mu - 2 \beta \bigg( B'(0) - \int_0^{v_c} dv B'\left( \int_k \tilde G(k) - G(k,v)\right)
\\
& - &(1-v_c) B'\left(\int_k \tilde G(k) - G(k,v_c)\right) \bigg)
\eea
which can be further rewritten using (\ref{e2vc}) and the definition of $\sigma(v)$ in (\ref{e1}) as:
\bea
\fl \mu_{\rm eff} &=& \mu - 2 \beta  B'(0) - \int_0^{v_c} dv \sigma(v) +v_c \sigma(v_c) +2\beta
B'\left( 2 T \int_k \frac{1}{\mu - t \Delta(k) + \Sigma_c} \right)
\\
\fl &=&\mu +\Sigma_c -\frac{2}{T}\left(B'(0)-
B'\left( 2 T \int_k \frac{1}{\mu - t \Delta(k) + \Sigma_c} \right)\right)
\eea

In the limit $T \to 0$, and recalling that $\Sigma_c = \mu_c - \mu$ we find
\be \label{mueffFRSB}
 \mu_{\rm eff} = \mu_c  + 4 B''(0) \int_k \frac{1}{\mu_c - t \Delta(k)}
\ee
This has the same form as the RS formula \eqref{muefRS}
where one replaces $\mu$ by the Larkin mass $\mu_c$, i.e.
it can be interpreted as the mass $\mu$ {\it freezing} at $\mu_c$, that is retaining for $\mu<\mu_c$ its critical value.

Let us now determine the lower edge of the Hessian. From \eqref{simple} we
obtain upon inserting \eqref{mueffFRSB}
\be
\lambda_e^- = \mu_{\rm eff} - \mu_c - 4 B''(0) \int_k \frac{1}{\mu_c - t \Delta(k)} = 0
\ee
Hence the lower edge of the Hessian remains frozen at zero within the FRSB phase
for all values of $\mu$.  For models with more than one edge, their positions can be found from the other roots of
the equation \eqref{Lark}  as discussed in the previous
Section. One should then insert them in \eqref{simple} and use
\eqref{mueffFRSB} {\it without inserting them in \eqref{mueffFRSB}}, the latter being defined in terms of the Larkin length $\mu_c$.

\subsubsection{1-step replica symmetry breaking phase}

We now study SRC potentials which exhibit the 1RSB solution. For the
continuum models, the 1RSB solution holds for $d \leq 2$ and
$\gamma \geq \gamma_c(d)=2/(2-d)$.\\

Let us give a brief account of the 1RSB parametrization and the ensuing procedure.
We start with introducing two parameters $\sigma_1$ and $\Sigma_c$ in terms of which we construct a $v_c\times v_c$ matrix $\sigma_d$
with entries $(\sigma_d)_{ab}=-\Sigma_c\delta_{ab}+\sigma_1$. The full $n\times n$ matrix $\sigma$ has $n/v_c$ identical diagonal blocks
$\sigma_d$, all entries being equal to the value $\sigma_0$ outside those blocks. The constraint $\Sigma_{b}\sigma_{ab}=0$ then  yields
the relation in the $n\to 0$ limit:
\be\label{constraint1RSB}
-\Sigma_c+v_c\sigma_1+(n-v_c)\sigma_0=0 \quad \Rightarrow \quad  v_c(\sigma_1-\sigma_0)=\Sigma_c
\ee
Inversion of the matrix $G^{-1}=\mu-t\Delta(k)+\sigma$ produces $n\times n$ matrix $G$ with the diagonal $v_c\times v_c$ blocks $G_d$ having entries $(G_d)_{ab}=(\tilde{G}-G_1\delta_{ab}+G_1$ and outside those blocks $G$ has identical entries $G_0$. The entries in the limit $n\to 0$ remembering (\ref{constraint1RSB}) are given by relations:
\be
G_0=\frac{\sigma_0}{(\mu-t\Delta(k))^2}, \quad \tilde{G}-G_0=\left(1-\frac{1}{v_c}\right)\frac{1}{\mu-t\Delta(k)+\Sigma_c} +\frac{1}{v_c}\frac{1}{\mu-t\Delta(k)}
\ee
and
\be
\tilde{G}-G_1=\frac{1}{\mu-t\Delta(k)+\Sigma_c},
\ee
which according to (\ref{repsp2chi}) leads to
\be\label{chi1RSB}
\chi_0=\frac{2 T}{v_c} \left( \int_k \frac1{\mu - t \Delta(k)} -  \int_k \frac{1-v_c}{\mu - t \Delta(k) + \Sigma_c}\right),
~~
 \chi_1=2 T  \int_k \frac1{\mu - t \Delta(k) + \Sigma_c}
\ee

To determine the equilibrium values of the parameters involved we rely upon the expression for the free energy $\Phi(T)$ associated with the model given \footnote{see Eq. (170)
in Section III.D p17 or the arXiv version.
in the paper \cite{LDMW} (up to an irrelevant constant).}
\[
\Phi(T) = \frac{1}{2 T} \left(v_c B(\chi_0) + (1-v_c) B(\chi_1) \right)
\]
\be
+ \,\,\frac{T}{2} \frac{1-v_c}{v_c}
 \int_k \left( \frac{\Sigma_c}{\mu - t \Delta(k) + \Sigma_c} - \log\left(\frac{\mu - t \Delta(k) + \Sigma_c}{\mu - t \Delta(k)}\right) \right)
\ee
Taking a derivative of the free energy w.r.t. $\Sigma_c$ leads to
\be
\Sigma_c = - 2 \frac{v_c}{T} ( B'(\chi_1)-B'(\chi_0))
\ee

Let us consider the $T=0$ limit. Denoting
\be\label{defQ1RSB}
v_c=v T \quad , \quad Q := \frac{2}{v} (I_1(\mu)-I_1(\mu+\Sigma_c))
\ee
in terms of the integrals defined in (\ref{In})
and noticing that in this limit $\chi_0\to Q$ and $B(\chi_1)/2T\to B'(0)I_1(\mu+\Sigma_c)$
one gets
\be
\Phi(0) = B'(0)I_1(\mu)+\frac{v}{2} \left( B(Q) - B(0) - Q B'(0)\right) - \frac{1}{2 v} F_\mu(\Sigma_c)
\ee
where we have defined
\be
F_\mu(x) = I_0(\mu+x)-I_0(\mu) - x I_1(\mu+x)
\ee
with $F_\mu(0)=F_\mu'(0)=0$ and $F_\mu''(0)=I_2(\mu)$.

Upon derivation of the zero-temperature free energy w.r.t. $\Sigma_c$ (cancelling the common factor $I_2$)
and $v$  one obtains the following system of equations:
\bea \label{sp1rsb}
&& \Sigma_c = 2 v (B'(Q)-B'(0)) \\
&& \frac{1}{v^2} F_\mu(\Sigma_c) = B(0) + Q B'(Q) - B(Q)
\eea
which should be augmented with the definition of $Q$ in (\ref{defQ1RSB}).
For small $Q>0$ we have $\Sigma_c \simeq 2 v B''(0) Q$ and substituting in  (\ref{defQ1RSB}) we then
find that the transition to the phase with nonzero value of $Q$  occurs at
$\mu=\mu_c$ determined by
\be \label{muc2}
1 = 4 B''(0) I_2(\mu_c)
\ee
which identifies with the definition of the Larkin mass, cf. (\ref{Lark}). \\

We can now give the formula for $\mu_{\rm eff}$ in the 1RSB phase. From
\eqref{mueffdef} we have, inserting the one-step RSB ansatz
\be
\mu_{\rm eff} = \mu - \frac{2}{T} (B'(0) - v_c B'(\chi_0) - (1-v_c) B'(\chi_1))
\ee
which in the limit $T \to 0$ yields
\be
\mu_{\rm eff} = \mu + 4 B''(0) I_{1}(\mu) + 2 v (B'(Q)-B'(0)- Q B''(0))
\ee
Recalling from \eqref{simple} that
\be
 \lambda_e= \mu_{\rm eff} -  4  B''(0) I_{1}(\mu_c) -\mu_c
\ee
we finally obtain, within the 1RSB phase
\be \label{edge1RSB}
 \lambda_e=  \mu  - \mu_c  + 4 B''(0) (I_1(\mu) - I_1(\mu_c))+ 2 v (B'(Q)-B'(0)- Q B''(0))
 \ee
providing the expression for the position of the lower spectral edge in the 1RSB phase.

We now expand below and near the transition: we insert $\Sigma_c$ from the
first equation in the second and third, which gives two coupled
equations for $Q$ and $v$. In these equations we insert, with $\delta>0$,
\bea
\mu=\mu_c(1-\delta) \quad , \quad Q=\sum_{n \geq 1} Q_n \delta^n \quad , \quad v=
\sum_{n \geq 0} v_n \delta^n
\eea
and solve order by order. It is convenient in the calculation to use that
$I_l(x)=(-1)^{l-1} I_0^{(l)}(x)/(l-1)!$ for $l \geq 1$. We give only the lowest
order
\be
v_0= - \frac{B'''(0)}{16 B''(0)^3 I_3(\mu_c)} \quad , \quad Q_1= - \frac{8 \mu_c B''(0)^2 I_3(\mu_c)}{B'''(0)}
\ee
recalling that $B'''(0)<0$. To this order one finds that the edge $\lambda_e$ vanishes
to order $O(\delta^2)$. To find the first non-vanishing order, $O(\delta^4)$ one
needs to calculate $v_1,Q_2,v_2,Q_3$ iteratively. Performing the calculation using the Mathematica software
we finally find, after some rearrangments using \eqref{muc2}, up to $O\left(\delta ^5\right)$ terms

\be \label{edge1rsb}
\lambda_e = \frac{\mu_c}{36 B^{(3)}(0)^4} \left(\frac{\mu_c I_3(\mu _c)}{I_2(\mu_c)}\right)^3
\left(B^{(4)}(0) B''(0) - 3 (1- \frac{I_2(\mu _c) I_4(\mu _c)}{2 I_3(\mu _c)^2}) B^{(3)}(0)^2
\right)^2  \delta ^4
\ee
This result is very general, for any discrete or continuum model. For the $d=0$ 'single particle' model, $I_l(\mu_c)=\mu_c^{l}$ for $l \geq 1$ and \eqref{edge1rsb} reduces exactly to the formula (76) obtained
in our previous work \cite{UsToy}.

We can now specify to the continuum model by setting $- t \Delta(k)= k^2$ (we set $t=1$ for simplicity) and recall that
$\int_k$ denotes $\int \frac{d^dk}{(2 \pi)^d}$. The equations
\eqref{sp1rsb} hold with
\be
F_\mu(\Sigma_c) = \mu^{d/2} {{\cal F}}(\frac{\Sigma_c}{\mu}) \quad , \quad {{\cal F}}(x) = \int \frac{d^d k}{(2 \pi)^d} ( \ln(1+ \frac{x}{1 + k^2} ) - \frac{x}{1 + k^2 + x})
\ee
\be
 Q = \frac{2}{v} \mu^{d/2-1}  {\cal G}(\frac{\Sigma_c}{\mu}) \quad , \quad {{\cal G}}(x) =
\int \frac{d^d k}{(2 \pi)^d} (\frac{1}{k^2+1} - \frac{1}{k^2+1+x})
\ee
Restricting our consideration to  $d<4$ one can see that ${\cal F}$ and ${\cal G}$ are defined by a UV convergent integral, so that
we can set the UV cutoff $k_{\max}$ to infinity.

We can check that the ratio entering in \eqref{edge1rsb} are
\bea
&& \frac{\mu_c I_3(\mu _c)}{I_2(\mu_c)} = \frac{4-d}{4}
\quad , \quad
 \frac{I_2(\mu _c) I_4(\mu _c)}{2 I_3(\mu _c)^2} = \frac{6-d}{3(4-d)}
\eea
hence we obtain for the continuum model in dimension $d$
\be
\lambda_e = \frac{\mu_c}{36 B^{(3)}(0)^4} \left(\frac{4-d}{4}\right)^3
\left(B^{(4)}(0) B''(0) - \frac{2 (3-d)}{4-d} B^{(3)}(0)^2
\right)^2  \delta ^4 +O\left(\delta ^5\right)
\ee

Let us study $d=1,2$ in more details. In $d=2$ for the continuum model we have
\be
{{\cal F}}(x) = \frac{1}{4 \pi} (x -  \log(1+x))
\quad , \quad {{\cal G}}(x) = \frac{1}{4 \pi} \log(1+x) \quad , \quad
I_1(\mu) - I_1(\mu_c) = \frac{1}{4 \pi}  \log \frac{ \mu_c}{\mu}
\ee
and the transition occurs at
\be
\mu_c = \frac{B''(0)}{\pi}
\ee
From the last equation in \eqref{sp1rsb} we find that $ \Sigma_c = \mu( e^{2 \pi Q v}-1 )$,
and substituting into the other two equations we obtain the system
\bea
&& \mu( e^{2 \pi Q v}-1 ) = 2 v (B'(Q)-B'(0)) \\
&& \frac{1}{2 \pi v} (B'(Q)-B'(0) - \mu \pi Q) = B(0) + Q B'(Q) - B(Q)
\eea
The second equation allows to obtain $v=v(Q)$ and reporting in the first one
it leads to an equation for $Q$.

Let us now specify to the exponential case $B(q)= e^{-c q}$ which we expect to be marginal at the boundary with full RSB. One finds that the
solution is remarkably simple. For $\mu< \mu_c=c^2/\pi$ it reads
\be
v = \frac{c}{2 \pi} \quad , \quad Q = \frac{1}{c} \log \frac{c^2}{\mu \pi}
\ee
Inserting into \eqref{edge1RSB} we find, for the exponential case, that the
edge of the spectrum of the Hessian is exactly at zero
\be
\lambda_e=0
\ee
for all $\mu \leq \mu_c$. This is the confirmation of the case being  marginal for
$d=2$, i.e. it can be obtained as a limiting case from the FRSB side.
It is interesting to note that its exact solution is also very simple.\\

Let us now consider the marginality  for $d=1$ when
\be
{{\cal F}}(x) = \frac{2 + x}{2 \sqrt{1+x}} - 1\quad , \quad {{\cal G}}(x) = \frac{x}{2( 1+ x + \sqrt{1+x}) }
\quad , \quad
I_1(\mu) - I_1(\mu_c) = \frac{1}{2 \sqrt{\mu}} - \frac{1}{2 \sqrt{\mu_c}}
\ee
and $I_2(\mu_c)= \frac{1}{4} \mu_c^{-3/2}$, leading to $\mu_c = (B''(0))^{2/3}$.
Let us choose
\be
B(q) = \frac{A}{c + q}
\ee
Then
\bea
\mu_c= \frac{(2 A)^{2/3}}{c^2} \quad , \quad  v_0 = (2 A)^{-1/3}  \quad , \quad Q_1 = \frac{c}{2}
\eea
We must now solve the equations
\bea
&& \Sigma_c = \frac{2 v A}{c^2} \left(1 - \frac{c^2}{(c+Q)^2}\right) \\
&& \frac{1}{v^2} \left(\frac{2 \mu + \Sigma_c}{2 \sqrt{\mu+\Sigma_c}} - \mu^{1/2}\right)
= \frac{A Q^2}{c (c+Q)^2} \\
&& Q v= \mu^{-1/2} - (\mu+\Sigma_c)^{-1/2}
\eea

It is convenient to introduce the following variables and parameters:
\[
\frac{c}{c+Q}=x, \quad \sqrt{\mu+\Sigma_c}=y, \quad Qv=z, \quad \Omega=\frac{2A}{c^3}
\]
in terms of which the above system takes the form
\bea
&& y^2=\mu+\Omega \,zx(1+x)  \\
&& \frac{(y-\sqrt{\mu})^2}{y}=\,\Omega\,z^2x^2\\
&& z=\frac{y-\sqrt{\mu}}{\sqrt{\mu}y}
\eea
Substituting the last of those equations to the second one and remembering that for $\Sigma_c>0$ we have $y>\sqrt{\mu}$ we see that
the second equation takes the form $y=\frac{\Omega}{\mu} x^2$ implying further that $z=\frac{1}{\sqrt{\mu}}-\frac{\mu}{\Omega x^2}$.
Substituting these relations into the first equation we see that it can be brought to the form
\[
\left(\frac{\Omega}{\mu} x^2\right)^2-\mu=\sqrt{\mu}\frac{1+x}{x}\left(\frac{\Omega}{\mu} x^2-\sqrt{\mu}\right) \quad \Leftrightarrow\quad
\left(\frac{\Omega}{\mu} x^2-\sqrt{\mu}\right)\left(\frac{\Omega}{\mu} x^2-\frac{\sqrt{\mu}}{x}\right)=0
\]
The first solution $x^2=\frac{\mu^{3/2}}{\Omega}$ is however not admissible since it corresponds to $y=\sqrt{\mu}$ and therefore to $\Sigma_c=0$. The only nontrivial solution as $\mu$ is decreased below $\mu_c$ is provided then by  the remaining root $x=\frac{\mu^{1/2}}{\Omega^{1/3}}$ and in the original variables finally yields the relations
\bea
v = (2 A)^{-1/3} \quad , \quad Q  + c = \frac{(2 A)^{1/3}}{\sqrt{\mu}}  \quad , \quad \mu + \Sigma =
\frac{(2 A)^{2/3}}{c^2}
\eea
Substituting in \eqref{edge1RSB} we again find $\lambda_e=0$, confirming marginality for this case.

\subsection{Spatial structure of the Green function, pinning and localization}

One of the interest in the manifold problem compared to the point ($d=0$) is the
rich internal space structure. The hierarchical construction of the Gibbs measure encoded in the RSB solution was
discussed in the context of the manifold in the Appendix of \cite{MP2} (see also discussion in \cite{TGPLDBragg2,LeD11}). In that picture the Gibbs measure is a superposition of Gaussians, with power law distribution of weights, each centered around distinct seed configurations ${\bf u}^\alpha(x)$, with fluctuations controled by an ``effective mass'' (each Fourier mode has its own decomposition into states).
The picture is either one-step (1RSB) or is hierarchically repeated (FRSB). It was shown that the closeby states (at $v=v_c$) correspond typically to
the scale of the Larkin length (with effective mass $\mu+\Sigma_c = \mu_c$),
while the large scale statistics (e.g. of
${\bf u}(x)- {\bf u}(0)$ at large $x$) is controled, throughout the glass phase,
by the small $[\sigma](v)$ bevahior (with effective mass $\mu+[\sigma](v)$).
Hence in the FRSB phase it is the small $v$ behavior, corresponding to distant states, which to the non-trivial roughness exponent. Here we study the Hessian at the global minimum and its
"soft modes" contain information the structure of the states (we saw in particular
that the gap is zero from the marginality condition).

We can thus now ask about the spatial structure contained in the averaged Green function. Let us examine again
the formula \eqref{Green}. If we choose $\lambda=\lambda_e^{-}$, i.e. at the lower edge we can write
\be
\overline{ G(x,y;\lambda=\lambda^{-}_e,{\bf u}_0)} = \int_k \frac{e^{i k (x-y)}}{\lambda^{-}_e -\mu_{\rm eff} + t \Delta(k) - 4 i p B''(0)}
= -  \int_k \frac{e^{i k (x-y)}}{- t \Delta(k) + \mu_c}
\ee
where we used \eqref{simple}. Hence we see that the averaged Green function decays exponentially
$\sim e^{-|x-y|/L_c}$,
with the characteristic length
given by the Larkin length $L_c$. For the continuum model with short-range elasticity and weak disorder,
$L_c \sim 1/\mu_c^{1/2}$.
This is very reminiscent of the result found numerically in
\cite{CaoRossoSoftModes} in the context of depinning. Remarkably however, here this property holds also for
$\mu>\mu_c$, i.e. in the RS phase.

Note that in the standard interpretation of the localization theory the decay rate of the
disorder-averaged Green's function in the bulk of spectrum defines the so-called {\it mean-free path} and generically has little to do with the true localization length. The situation at the spectral edge may however be different as, in contrast to the bulk, in that region of spectrum the Green function is not expected to show fast oscillations with random phase in every disorder realization, whose averaging gives rise to the decaying mean.
We therefore expect that the decay rate at the edge may have relation to the localization properties of the lowest eigenmode of the Hessian.

\section{Conclusion}
\label{sec:conclusion}

In this paper we have extended our previous work on the spectrum of the Hessian matrix at the global minimum
of a high dimensional random potential, to the case of many points coupled by an elastic matrix. This is of
interest in several contexts, in particular for disordered elastic systems pinned in a random environment.
We have calculated the averaged Green function and its imaginary part, the spectral density, of the Hessian matrix.
Technically this was achieved using a saddle point method and two sets of replica, one to express the Green function, the
second to impose the constraint of global minimum. The latter requires a replica symmetry breaking solution
for the saddle point equations, either of the 1 step kind (1RSB) or with full replica symmetry breaking (FRSB).
We have derived the criterion according to which one has the former or the latter, which generalizes the concept of
short range (leading to 1RSB) or long range (FRSB) disorder to the case of the elastic manifold.

The main difference with the case of the particle $d=0$ in a random potential is that the spectral density of
the Hessian is not a semi-circle anymore.
We have calculated its form in a number of examples and
obtain the values of the edges.
We have shown how it can evolve from a many band to a single band as the disorder is increased. In all generic cases however it retains a semi-circle shape near its edges.  Especially complete and explicit characterisation of the arising spectral density has been achieved in the 1D continuous system of infinite length.

Concerning the position of the lower edge, we have shown that qualitatively the scenario found for the
particle remains valid for the manifolds. For short range disorder cases and $\mu>\mu_c$ the Hessian spectrum is
gapped away from zero. At $\mu=\mu_c$ the gap vanishes, i.e. the lowest eigenvalue is zero.
For $\mu < \mu_c$ the saddle-point solution is 1RSB and we find that the gap is non zero and vanishes as
$\sim (\mu_c-\mu)^4$ near the transition. For long range disorder cases we find that the
gap vanishes identically for $\mu\le \mu_c$, reflecting the marginality of the FRSB solution.
 We also identified and studied the cases of marginally correlated disorder in $d=1$ and $d=2$
which can be of separate interest.

A new feature which emerges in the study of the manifold is the information about the internal spatial dependence of the averaged Green function. We found that near the edge it decays over a length scale identical to the so-called Larkin length,
related to $\mu_c$, which plays a central role in the theory of pinning. Below the Larkin scale the system responds elastically, while above the Larkin scale, metastability sets in leading to glassy non linear response. Our result in the high embedding dimension limit, are reminiscent to what was found
in numerical simulations for elastic strings at the depinning transition, where the localization length of the low lying modes of the Hessian was
found to be equal to the Larkin length.

Many questions remain. One is to understand the statistics of the lowest eigenvalues. Clearly it cannot
be of the Tracy Widom type since it is bounded by zero. The question of its universality remains open.
One possible way to tackle this difficult problem is to study the large deviations for the minimal eigenvalue.
 Another interesting problem is to generalize counting analysis of the minima and saddle-points from the particle case $d=0$ \cite{Fyo04,FyoWil07,FyoNad12} to the present manifold model with $d\ge 1$.  Work is presently in progress in those directions.

Finally, the most interesting but very challenging problem is to study the present model by taking limits
 $N\to \infty$ and $L\to \infty$ in a coordinated way, and scaling the coupling $t$ accordingly to enter the regime when Anderson localization effects in Hessian spectrum should be dominant. It remains to be seen if field-theoretical/supersymmetric methods which proved to be instrumental in getting insights into spectra and eigenvectors of matrices of banded type \cite{FM1991,Shch2_2018} could be used successfully in the present problem.

\bigskip

{\bf Acknowledgments:} YVF is grateful to Mira Shamis for a helpful discussion about the 'deformed semicircle'  in the context of Wegner orbital model.  This research was initiated during YVF visit to Paris supported by
the Philippe Meyer Institute for Theoretical Physics, which is gratefully acknowledged.
The research at King's (YVF) was supported by  EPSRC grant  EP/N009436/1 "The many faces of random characteristic polynomials".
PLD acknowledges support from ANR grant ANR-17-CE30-0027-01 RaMa-TraF.

\appendix

\section{analysis of $\delta L[Q,\sigma,P,\tau,R,\eta]$}

We give here the last piece of the replicated action, omitted in the text.
\bea\label{lastpiece}
\fl && \delta L[Q,\sigma,P,\tau,R,\eta] = \frac{1}{2} \Tr \ln \begin{pmatrix}
(\mu \mathbf{1} - t \Delta) \mathbf{1}_n - \sigma \mathbf{1}   & \eta \\
\eta^T &  (\mu \mathbf{1} - t \Delta) \mathbf{1}_m - \tau \mathbf{1} )
\end{pmatrix}
\\
\fl && -  \frac{1}{2} \Tr \ln ((\mu \mathbf{1} - t \Delta) \mathbf{1}_m - \tau \mathbf{1} )
 - \frac{1}{2} \Tr \ln ((\mu \mathbf{1} - t \Delta) \mathbf{1}_n - \sigma \mathbf{1} )
\\
\fl && + 2 i\beta \sum_{a=1}^n B''\left(\frac{Q_{aa}+Q_{a1}-2Q_{11}}{2}\right)
\left( (R R^T)_{aa} + (R R^T)_{11} - 2 (R R^T)_{a1} \right) \\
\fl && - \frac{1}{2} \sum_{x} \tr (\eta(x) R(x))
\eea
At the saddle point $R=0$ hence it vanishes. The main argument for
that is very similar to the discussion in \cite{UsToy}.


\end{document}